\documentclass[aps,preprint,prb]{revtex4}
\usepackage{graphicx}
\usepackage{epsfig}
\usepackage{amsmath}
\usepackage{graphics}
\usepackage{latexsym}
\begin{document}

\author{B.\ M.\ Mognetti, L.\ Yelash, P.\ Virnau, W.\ Paul, K.\ Binder}
 \altaffiliation[]{} 
 \email{kurt.binder@uni-mainz.de}
\affiliation{Institut f\"ur Physik, Johannes
Gutenberg-Universit\"at, Mainz,
 Staudinger Weg 7, D-55099 Mainz, Germany %
}%
\author{M.\ M\"uller }
\affiliation{Institut f\"ur Theoretische Physik,
Georg-August-Universit\"at G\"ottingen, Friedrich-Hund-Platz 1,
D-37077 G\"ottingen, Germany }
\author{L. G. MacDowell}
\affiliation{
 Dpto. de Quimica Fisica, Facultad de Cc.
Quimicas, Universidad Complutense,
28040 Madrid, Spain}

\title{Efficient prediction of thermodynamic properties of quadrupolar fluids  from simulation of a coarse-grained model: The case of carbon dioxide}

\date{\today}

\begin{abstract}
Monte Carlo simulations are presented for a coarse-grained model
of real quadrupolar fluids. Molecules are represented by
particles interacting with Lennard-Jones forces plus the
thermally averaged quadrupole-quadrupole interaction. The
properties discussed include the vapor-liquid coexistence curve,
the vapor pressure along coexistence, and the surface
tension. The full isotherms  are also accessible over a wide range of
temperatures and densities. It is shown that the
critical parameters (critical temperature, density, and pressure)
depend almost linearly on a quadrupolar
 parameter $q=Q^{*4} /T^*$,  $Q^*$
is the reduced quadrupole moment of the molecule and $T^*$ 
the reduced temperature.

The model can be applied to a variety of small quadrupolar
molecules. We focus on carbon dioxide as a test case, but consider
nitrogen and benzene, too. Experimental critical temperature, density
and quadrupolar moment are sufficient to fix the parameters of the model.
The resulting agreement with experiments is excellent and marks
a significant improvement over approaches which
neglect quadrupolar effects.
 The same coarse-grained model
was also applied in the framework of 
Perturbation Theory  (PT) in the Mean Spherical
Approximation (MSA). As expected,  the latter deviates from
the Monte Carlo results in the critical region, but is reasonably
accurate at lower temperatures.

\end{abstract}

\pacs{Valid PACS appear here}
\maketitle

\section{\label{sec:level12} Introduction}
 
Solvents play an essential role
in  the design and processing of many molecular materials (e.g.,
oligomers, polymers, etc). 
In comparison to a melt, the molecular mobility of dissolved
substances increases considerably in solution. Not only
the flow properties can be controlled easily in solution, but also the phase
behavior (and hence the morphology) of the dissolved materials,
e.g., by changing the thermodynamic state conditions like temperature,
pressure, and concentration.

A particularly important solvent is supercritical carbon dioxide,
because the material is inexpensive, nonpoisonous, not reactive, and
thermally stable. Hence, its application as a solvent is
widespread. \cite{KB-93,KLS-94,KM-05} However, the phase behavior of
polymer-solvent systems or other binary fluid mixtures 
is rather complex in general. When the thermodynamic control parameters
temperature $T$, pressure $p$ and solute molar fraction $x$ are
varied, various liquid-vapor and fluid-fluid phase equilibria
occur, and  many different types of (rather complicated) phase
diagrams can be observed \cite{RS-82,KS-80}. Even for simple binary fluid
mixtures, e.g., carbon dioxide plus short alkanes such as
hexadecane, the phase diagram is only known rather incompletely
from experiment. \cite{SAHHF-67,AMK-86} These uncertainties also hamper the
judgment of the accuracy of the theoretical modeling of such
systems. \cite{VMMB-04,VMMB-02,BMVM-05,IKHUA-95} In fact, due to the large control
parameter-space that needs to be explored, comprehensive
experimental work would be very cumbersome, and a modeling
approach seems to be the method of choice. However, the large number of
states $(T,p,x)$ that need to be simulated and the complexity of
the systems renders a fully chemically realistic all-atom simulation
practically impossible. Thus, the construction of a
suitable coarse-grained model for such systems containing
polymers (or oligomers, respectively) is very desirable. While there
is a rich literature on the construction of coarse-grained models
for (flexible) polymers, \cite{PBKH-91,TKBBH-98,BBDGHKMMMPSST-00,KM-01,M-02,AK-03,BPSS-04,V-07}
comparatively little attention has been paid to the question on how a
coarse-grained solvent molecule such as CO$_2$ should be
described. Iwai et al.\ \cite{IKHUA-95} and  Virnau et al. \cite{VMMB-04,VMMB-02,BMVM-05} simply used
particles interacting with simple Lennard-Jones forces among
themselves and with the beads of the bead-spring chain that
represents effective subunits of the polymer. While
particles interacting with Lennard-Jones potentials describe noble
gases such as liquid argon or neon rather well, it is clear that
a ``Lennard-Jonesium'' is a somewhat unsatisfactory description of a
carbon dioxide molecule. While considerable attention has been paid
to atomistic
models of CO$_2$, 
\cite{MSM-81,BMA-84,BA-85,ZR-89,EK-89,GLC-90,PG-93,HY-95,VHRPM-00,ZD-05,BHS-07,BBV-00,BSJJSKWR-99,SVHF-01,VSH-01,GLVMR-94} we are not
aware of comprehensive systematic studies of coarse-grained models for this
molecular fluid. With respect to the atomistic models for CO$_2$,
we note that there is no consensus in the literature on a unique
form of the interaction potential and its parameters.  
Starting from the Murthy-Singer-McDonald (MSM) model, \cite{MSM-81} several
potentials have been proposed (for a recent comparison 
see Ref.\ \onlinecite{ZD-05}). In Ref.\ \onlinecite{HY-95}, 
two variants of the Elementary Physical Model (EPM) force field were suggested, 
that yielded critical temperatures of $T_c=313.4 \, \pm
0.7$ K and $312.8\, \pm 3.0$ K, respectively, while the experimental
value is $T_c=304.2$ K. \cite{HCP-85} 
In view of this 3\% discrepancy
between the atomistic models and the experiment, it was suggested
\cite{HY-95} to use  the experimental critical
temperature and rescale the energy parameters of the model to
reproduce the correct value of the critical temperature (EPM2).
In fact, Virnau et al.\ \cite{VMMB-04,VMMB-02,BMVM-05}, using a simple
``Lennard-Jonesium'' to model CO$_2$, fixed the Lennard-Jones
parameters to match both the critical temperature $T_c$ and 
the critical density
$\rho_c$. Since the atomistic models underestimate the
critical density (yielding \cite{HY-95} \, $453.7 \, \pm 4.3 $ kg$/$m$^3$
or $449 \pm 16 $  kg$/$m$^3$ instead of the experimental value
\cite{HCP-85} $\rho_c=468.0 $  kg$/$m$^3$), they also require a corresponding
rescaling of the interaction range parameters. Hence, EPM2
 needs the same input from the experimental critical data as
the coarse-grained model of Virnau et al.\ \cite{VMMB-04,VMMB-02,BMVM-05}. 
For the resulting model, the coexistence
densities predicted for the liquid branch in the
temperature region $230$K $\leq T \leq 280$K deviate distinctly less
from the experimental results \cite{HCP-85} 
than the  
corresponding
results of the coarse-grained model. \cite{VMMB-04,BMVM-05}

As indicated above, the main interest for obtaining an
accurate coarse-grained model for CO$_2$ is 
its potential application in  multicomponent systems,
e.g., polymer solutions in which CO$_2$ acts as a solvent. \cite{VMMB-04,BMVM-05}
For such systems also many attempts 
were undertaken to derive approximate analytical equations of
states (e.g., Refs.\ \onlinecite{GB-02,PWS-03}) and it is, of course, also highly
desirable to validate such equation of state theories by simulations.
However, the coarse-grained model for CO$_2$ of Virnau et al.\
\cite{VMMB-04,BMVM-05}, when combined with a suitable coarse-grained model for
the alkanes, required rather large deviations
from the simple Lorentz-Berthelot mixing rules 
to account  for the
available experimental data. \cite{SAHHF-67,AMK-86} 
Most likely, the somewhat oversimplified CO$_2$ model
is responsible for most of these deviations. Approximating
CO$_2$ as a Lennard-Jones particle without
considering its rather large quadrupolar moment ($|Q|=4.3$ D$\mathrm{ \AA}$)
is probably not
sufficient -- the unit D (Debye)  equals  $10^{-18}$ in CGS units which are adopted 
throughout the manuscript.

In the present paper we explore a slightly more involved
coarse-grained model for CO$_2$.\cite{SRN-74,MG-03}
  The molecule is still described as
a Lennard-Jones particle, but we also include
 the experimentally known quadrupole moment as an
input parameter, together with critical temperature
and critical density. A precondition for the usefulness of
coarse-grained models is that simulation codes execute very fast.
The angular-dependent quadrupole-quadrupole interaction requires
significant computational resources which would be a serious
drawback to such a model.
However, compared to the
Lennard-Jones forces, the quadrupolar interaction is still a rather
weak perturbation. Therefore, we apply one further approximation:
\cite{SRN-74}  the angular dependence is averaged over in a second
order thermodynamic perturbation calculation.
Thus,  an effective isotropic
potential is obtained. Rather encouraging results using such an
approximation have been reported in the literature. \cite{MG-03}
M\"uller and Gelb \cite{MG-03} estimate coexistence
curves from non-equilibrium molecular dynamics (NEMD) simulations
of temperature quenches from the one-phase region into the
two-phase region, where one then waits until the system has phase
separated into the two coexisting phases. \cite{GM-02,AM-03} In this manuscript
we  apply
grand-canonical Monte Carlo methods, \cite{B-97,LB-05,W-97} combined with
a finite size scaling \cite{W-97,F-71,B-92} analysis.
This allows us to
locate precisely the critical point of the model. Note that 
a direct estimation of the critical
point from the simulation is difficult
if either
Gibbs ensemble techniques \cite{P-92,P-87} or the temperature
quench technique \cite{MG-03,GM-02} are applied. In these 
cases, one  relies on a fit
of the coexistence data to a suitable power law extrapolation. With the
present techniques one can obtain the critical properties very 
accurately. This precision is required because
the
critical properties are used to gauge the Lennard-Jones
parameters of the model.

In Sec.\ II, we  give a more detailed
description of our model and simulation
techniques. 
Sec. III describes  our results for carbon dioxide and
compares them to previous approaches. Sec.\ IV discusses the
 application of the model to other quadrupolar fluids, namely
nitrogen and benzene.  Sec.\ V
describes the application of first order  
perturbation theory in the mean spherical approximation (PT-MSA)
to precisely the same model which was used in the simulation, thus
allowing a meaningful comparison. Finally, Sec. VI 
concludes the discussion and gives an outlook on future work.

\section{\label{sec:leve2}Model and Simulation Technique}
\subsection{\label{sec:leve2IIA.}Choice of Model}

Our model system consists of neutral spherical particles which carry a
quadrupolar moment $Q$ and interact with each other both via the
Lennard-Jones potential

\begin{equation} \label{eq1}
U^{LJ}_{ij}=4 \varepsilon\Bigg[\Big({\sigma \over r_{ij}} \Big)^{12} - 
\Big( {\sigma\over r_{ij}}\Big)^6
\Bigg]
\end{equation}

and the quadrupole-quadrupole interaction \cite{GG-84}

\begin{equation} \label{eq2}
U^{QQ}_{ij} =\frac{3 Q^2}{4 r^5_{ij}} \, f^{QQ} (\theta_i,
\theta_j, \phi_{ij}) \, .
\end{equation}

The angle-dependent part is given by:

\begin{eqnarray} \label{eq3}
f^{QQ} = 1-5 \, \cos^2 \theta_i - 5 \cos^2 \theta_j +17 \cos^2
\theta_i \cos^2 \theta_j \nonumber\\
+2 \sin^2 \theta_i \sin^{2}\theta_j \cos^2 (\phi_i-\phi_j)\nonumber\\
-16 \sin \theta_i \cos \theta_i \sin\theta_j \cos \theta_j \cos
(\phi_i-\phi_j) \quad .
\end{eqnarray}

In Eqs.~(\ref{eq1},~\ref{eq2}), $r_{ij}=|\vec{r}_i-\vec{r}_j|$ is
the distance between molecules at sites $\vec{r}_i$, $\vec{r}_j$,
while $(\theta_i, \phi_i)$ are the polar angles characterizing the
mutual orientations of the (linear) molecules 
(
$\theta_i$ are the angles between the axis joining
the two molecules and the quadrupole vectors of the molecules;
$\phi_i$ are the rotational orientations of the quadrupole vectors 
relative to the joining axis).
In Eq.~(\ref{eq1}), $\varepsilon$ and $\sigma$ set the scales of
energy and distance for the Lennard-Jones (LJ) interaction, 
respectively.

The  angular-dependent part of the potential
(Eqs.~(\ref{eq2}),~(\ref{eq3}))  slows
down the speed of the algorithm considerably.
Therefore, following Ref.\ \onlinecite{SRN-74}, 
we average  over the angles of the quadrupolar potential 
to create an effective isotropic representation.
More precisely, one expands the
Boltzmann factor $\exp (-\beta U ^{QQ}_{ij})$, ($\beta
=(k_BT)^{-1}$), in a Taylor series to second order in $\beta$.  
 After taking averages over the angles,  the following temperature-dependent
isotropic potential  is obtained:

\begin{equation} \label{eq4}
U^{IQQ}_{ij} =- \frac{7 \beta}{5} \, \frac{Q^4}{r^{10}_{ij}} \quad .
\end{equation}

For the potentials of Eq.~(\ref{eq1}) to Eq.~(\ref{eq4}) 
one can employ the standard procedure \cite{B-97,LB-05} of
cutting and shifting to zero at a cutoff distance 
$r_{ij}=r_c=2\sqrt[6]{2} \sigma$ typically applied to 
Lennard-Jones systems.
The total
potential then reads 

\begin{eqnarray} \label{eq5}
U(r_{ij})=  \left \{
\begin{array}{ll}
4 \varepsilon \Bigg[\Big({\sigma \over r_{ij}} \Big)^{12} - 
\Big( {\sigma\over r_{ij}}\Big)^6
-\frac{7}{20} q \Big({\sigma\over r_{ij}}\Big)^{10}
+S \Bigg] \, ,
\quad r \leq r_c \, .\\
0 \quad , \quad r\geq r_c\, .
\end{array} \right .
\end{eqnarray}
The reduced quadrupolar interaction parameter is defined as
\begin{eqnarray} \label{eq6}
q = {Q^4 \over \varepsilon \sigma^{10} k_BT } 
= q_c {T_c\over T} \, , &\qquad \qquad & q_c = q(T_c)
\end{eqnarray}
$q_c$ and $T_c$ are the values of the reduced quadrupole parameter
and temperature at the critical point.
$S$ shifts the cut potential to zero at $r_{ij}= r_c$,
so that $U(r_{ij})$ is  continuous everywhere: 
\begin{equation} \label{eq7}
S={ 127 \over 16384} + {7\over 5} {q\over 256} \quad .
\end{equation}
Note that Eq.\ (\ref{eq6}) is given in CGS units. In SI units,
there is
 an additional factor $(4 \pi \varepsilon_0)^{-2}$.
 It is
clear that  $U(r_{ij})$  is explicitly
temperature-dependent
because  $q$ and $S$  are temperature-dependent.
 Hence, special care needs to be exerted when
temperature derivatives are taken. For instance,  the fluctuation
relation linking the specific heat to the fluctuations of the
potential energy no longer holds for Eq.~(\ref{eq5}). We also
note that Eq.~(\ref{eq5}) differs from the potential obtained when
one  cuts off Eqs.~(\ref{eq1},\ref{eq2}) at $r_{ij}=r_c$. Indeed,
 continuity of $U^{QQ}_{ij}$ would require an
orientation-dependent shift of the potential.
It is also well-known \cite{FS-02} that the relation between the
critical temperature of a fluid and the energy scale $\varepsilon$
of the LJ interaction depends rather strongly on the cutoff $r_c$.
Our choice of a rather small value for the cutoff is mainly
motivated by the desire to have a very fast simulation algorithm,
but larger cutoffs will lead to very similar results.
As we will demonstrate later, differences in the phase diagram
almost disappear when simulation data are rescaled to match the 
experimental critical point for different $r_c$.
A further motivation for this choice of the cutoff
is that for $q=0$ our model reduces to that of Refs.\ \onlinecite{VMMB-04,BMVM-05}.

Our strategy  will be to compute the  critical
temperature $T_c(q_c)$ and  the critical density $\rho_c (q_c)$ from
the simulation, using the potential from Eq.~(\ref{eq5}).
Following previous work
\cite{VMMB-04,BMVM-05},  $\varepsilon$ and $\sigma$ 
are determined by the condition
that these critical parameters  match precisely their experimental
counterparts. In the following,  $T^*$ and $\rho^*$ 
will refer to 
temperatures and densities (and
other quantities that will be introduced with "$^*$")
expressed in units of $\varepsilon(q_c)$, $\sigma(q_c)$ 
and $M_\mathrm{Mol} $,
the molar mass of the fluid. 
We need to consider that the parameter
$q_c$, Eq.~(\ref{eq6}), depends itself on $\varepsilon$ and
$\sigma$, and not only on the (given) experimental value for $Q$. This
difficulty is related to the quadrupolar interaction in
Eq.~(\ref{eq5}), which shifts the critical point in the ($T, \rho$)
plane relative to its position for $Q=0$. Even if one is only
interested in a single choice of $Q$,  a simulation of a single
model system (i.e., one choice of $\varepsilon, \sigma$ and $Q$
or $q$, respectively) is never sufficient to deal with this
problem.
However, this puzzle can be solved by determining the critical
lines $T^*_c(q_c)$ and $\rho^*_c(q_c)$ as a function
of the (dimensionless) parameter $q_c$. Fig.~\ref{fig1}
shows the results of this calculation and demonstrates that both
$T^*_c(q_c)/T^*_c(0)$ and $\rho^*_c(q_c)/\rho^*_c(0)$ are very smooth
functions of $q_c$. These curves are almost linear,
so recording a few (altogether 9)
choices of nonzero $q_c$ was  sufficient to obtain 
good accuracy. In the range $0
\leq q_c \leq 0.5$, the critical temperature increases
 by almost 30\% while the critical density  increases
by about 10\%.

Having determined  $T^*_c(q_c)$ and $\rho^*_c(q_c)$,
 one can compute easily $\varepsilon(q_c)$ and
$\sigma(q_c)$  such that the model corresponds to a specific
experimental system $T_{c, \exp}$ and $\rho_{c, \exp}$. 
 Eq.~(\ref{eq6}) must hold together with

\begin{equation} \label{eq8}
\varepsilon(q_c) = k_BT _{c, \exp}/T^*_c(q_c) \, , \quad\quad \sigma^3 (q_c)
=\Big[\frac{\rho^*_c(q_c) M_{\rm Mol}}{\rho_{c, \exp} N_A} \Big] \quad
.
\end{equation}

Here $M_{\rm Mol}$ is the  molar mass of the simple molecule and
$N_A$ Avogadro's number. These equations are solved by a simple
iteration procedure, using the following fit functions
representing the data of Fig.~\ref{fig1},

\begin{subequations}
\begin{eqnarray}
\label{eq9a} T^*_c(q_c)/T^*_c(0) &=&1+ 0.46111 \, q_c + 0.17571\, q_c^2,\\
\label{eq9b}
 \rho^*_c(q_c) / \rho^*_c(0) &=& 1 + 0.19298\, q_c \quad,
\end{eqnarray}
\end{subequations}

\noindent 
where T$_c^{*}(q_c=0)=0.99821$ and $\rho_c^{*}(q_c=0)=0.32276$.
Appendix \ref{appB} explains in detail how simulation
parameters are derived from experimental data.
We note that the limiting factor for the accuracy of our procedure
is not at all the limited accuracy of Eqs.
~(\ref{eq9a},~\ref{eq9b}), but rather the uncertainty with which
the physical quadrupole moment $Q$ of the molecule, needed as
an input to Eq.~(\ref{eq6}), is known. Considering CO$_2$ as an
example, we take $Q=(-4.3 \pm 0.2)$ D$\mathrm{ \AA}$. 
However, since $Q$ is raised to the fourth power in Eq.~(\ref{eq6}),
 the 5\% uncertainty in $Q$ becomes a 30\%
uncertainty in the reduced simulation parameter $q$.
For $Q=-4.3$ D$\mathrm{ \AA}$, 
we obtain

\begin{equation} \label{eq10}
q_c =0.387, \quad \varepsilon=3.491 \times 10 ^{-21} J, \quad
\sigma=3.785~\mathrm{\AA}
\end{equation}

The uncertainty in $Q$  would actually allow for a range 
  $0.32 < q_c< 0.47$ with corresponding changes
of $\varepsilon$ and $\sigma$.
In view of these uncertainties, one could not hope for a perfect
agreement between the simulation results (for other quantities
rather than $\rho_c$ and $T_c$) and experiment, even if the form
of the coarse-grained potential, Eqs.~(\ref{eq5}-\ref{eq7}), were
perfectly accurate.

Already at this point, we note that nothing in the model
(Eqs.~(\ref{eq5})-(\ref{eq7})) is specific to CO$_2$. Hence,
Fig.~\ref{fig1} (or Eqs.~(\ref{eq9a},~\ref{eq9b}), respectively)
can be used for
modeling other quadrupolar
fluids, too. This fact will be taken up in Sec. IV and
Appendix \ref{appB}.
We also note that 
$\varepsilon$ and $\sigma$  are independent of the state
of the system once they are fixed.  $q$, however,
is given by $q=q_c \cdot T_c/T$
 (according to Eq.~(\ref{eq6})), 
 which needs to be
considered
 when  coexistence curve and
interfacial tension are calculated.

\subsection{\label{sec:leve2IIB.} Comments on the Simulation
Technique}

In this section we  comment briefly on the Monte Carlo simulation techniques
which are required for the 
computation of Fig.~\ref{fig1} and  other
physical properties. As in previous work, \cite{VMMB-04,BMVM-05} extensive
simulations were undertaken
in the $\mu VT$ ensemble, where the box volume
$V=L^3$, the chemical potential $\mu$ of the particles and the
temperature are fixed. The particle number fluctuates,
since the elementary Monte Carlo move consists of  random insertions or
deletions of particles. Thus,
long wavelength fluctuations of the density are  
equilibrated easily. In contrast, Molecular Dynamics or canonical
ensemble Monte Carlo methods that conserve the particle number in
the system suffer from a slow equilibration of long wavelength
density fluctuations (``hydrodynamic slowing
down'' \cite{LB-05}). The temperature quench simulations encounter
the additional difficulty that vapor-liquid interfaces extending
throughout the simulation box are formed. Such interfaces are
notoriously slowly relaxing and strongly fluctuating objects and
thus avoided in Gibbs ensemble techniques. \cite{P-92,P-87} 

For the sake of efficiency, histogram extrapolation techniques are
used. In a typical MC run, the particle number $n$ and the total
energy $E$ are recorded at regular intervals. The resulting distribution
$P_{\mu, T} (n, E)$ can then be extrapolated to neighboring values 
of  $\mu'$ and $T'$ using the following expression \cite{FS-88}

\begin{equation} \label{eq11}
P_{\mu',T'} (n,E) =\frac{1}{\mathcal{N}} \, P_{\mu, T} (n, E) \exp
\Big[\Big(\frac{\mu'}{T'} - \frac{\mu}{T}\Big)n -
\Big(\frac{1}{T'}-\frac{1}{T}\Big)E\Big] \,
\end{equation}
with ${\mathcal{N}}$ being a normalization constant.
Here, we have assumed that $q_c$ remains constant. 
Extrapolations at constant $Q$ would require an additional reweighting
factor related to the temperature-dependence of the potential
(Eq.~\ref{eq5}). Of course, Eq.~(\ref{eq11}) is only accurate  when
$P_{\mu, T}(n,E)$ and $P_{\mu', T'} (n, E)$ overlap strongly.
Nevertheless, reweighting is very useful for $\mu$ near
$\mu_{\rm coex} (T)$, where two-phase coexistence between vapor
and liquid occurs. In this region, $P_{\mu, T} (n)= \int d
EP_{\mu, T} (n,E)$ has a two-peak structure: one peak occurs at
$\rho^{(1)}_{\rm coex}\approx n/V $, the vapor density at
coexistence, the other peak at $ \rho^{(2)}_{\rm
coex}\approx n/V$, the liquid density at coexistence. For
$\mu=\mu_{\rm coex}(T)$,  the areas underneath
both peaks are equal (``equal area rule'' \cite{BL-84,BK-90}), but
unfortunately $\mu_{\rm coex}(T)$ is not known  beforehand.
However, if one has $P_{\rm \mu, T}(n)$, for some $\mu$ close
enough to $\mu_{\rm coex} (T)$, one can try to reweight
the data according to Eq.~(\ref{eq11}) with no additional simulation
effort. In this way, the coexistence curve can be located
precisely. The corresponding pressure  is  computed from the
virial equation. All these procedures have already been applied in
previous work for $q_c=0$. For  more details the reader is
referred to Refs.\ \onlinecite{VMMB-04,BMVM-05}.

Following a path along $\mu=\mu_{\rm coex} (T)$ in the $(\mu, T)$
plane and recording moments of the density distribution, we
calculate $2^\mathrm{nd}$ and  $4^\mathrm{th}$ order cumulants

\begin{equation} \label{eq12}
U_2=\langle M^2 \rangle / \langle |M| \rangle ^2 \quad , \quad
U_4= \langle M^4 \rangle / \langle M^2 \rangle^2
\quad , \quad M\equiv\rho- \langle \rho \rangle \, .
\end{equation}

Reasonably accurate estimates for $T_c$ can be obtained from the
intersection point of either $U_2(T)$ or $U_4(T)$  
for different $L$. The justification of this simple recipe follows
from the theory of finite size scaling. \cite{B-97,LB-05,W-97,F-71,B-92}
Fig.~\ref{fig2}  shows that 
 $T_c$ can be determined
 with a relative accuracy of about 3/10$^3$ 
with moderate computational effort. The lack of perfect
intersections in the size range $9 \sigma \leq L \leq 13.5 \sigma$
indicates that the asymptotic region of finite size scaling has
not been reached yet, and corrections to finite size scaling are still
present. However, the estimate $k_BT_c/\varepsilon=1.152 \pm 0.003$ 
is clearly  accurate enough for our present purposes. Note that the simple
analysis presented in Fig.~\ref{fig2}  ignores ``field
mixing''-effects \cite{W-97} between density and energy per particle. Of
course, for a high precision study of
critical exponents and critical amplitudes, more sophisticated
finite size scaling methods are available, \cite{KFL-03} but this
is beyond the scope of the present investigation.

For temperatures distinctly below $T_c$, the double-peak
distribution $P_{\mu, T} (n)$ 
 exhibits a deep minimum for densities $\rho$
in between the two coexisting phases $\rho^{(1)}_{\rm coex}$[vapor]
 and $\rho^{(2)}_{\rm coex}$[liquid].
\cite{B-82} Consequently, a system starting with a
low vapor-like density would hardly ever make the transition to
the liquid-like state or vice versa. Hence,
the relative weights of the two phases would not be sampled
correctly. This
difficulty is overcome by biased sampling methods that ``drive''
the system through the coexistence region such as
``multicanonical sampling'', \cite{BN-92} ''Wang-Landau-sampling''
\cite{WL-01} or ``successive umbrella sampling'' \cite{VM-04} which
has been used in this work. 
In the simplest implementation, 
 the algorithm is constrained to sample
configurations with only two particles  $n\in$ (0, 1) in the beginning, and 
 (1, 2)
$\cdots$ (n-1,n) later on, spanning the relevant range of densities.
The probability distribution can then be calculated recursively:
\begin{equation}
{P(n)\over P(0)} = H_{1,0} H_{2,1} \cdots 
H_{n,n-1} \quad ,
\end{equation}
with $H_{j,j-1}$ being  the frequency of occurrence of the $j^\mathrm{th}$
particle over the frequency of occurrence of the $(j-1)^\mathrm{th}$
particle in the sampling of the ($j-1$, $j$) window. 
For a more detailed
description of this method and its extension we refer to
 Virnau et al. \cite{VMMB-04,BMVM-05,VM-04}. 
Biased  grand canonical methods
 have the additional advantage that the minimum in $P_{\mu,T}(n)$
at densities near the density of the rectilinear diameter
$\rho_d(T)$

\begin{equation} \label{eq13}
\rho_d(T)=(\rho^{(1)}_{\rm coex} + \rho ^{(2)}_{\rm coex})/2 \,
\end{equation}

is also sampled rather accurately. This minimum
\cite{B-97,LB-05,W-97,B-82} 
corresponds 
 to a free energy barrier $\Delta F
\approx 2 \gamma (T) L^2$ which arises from the formation of two (planar)
vapor-liquid interfaces of area $L^2$, each connected with itself via
 periodic boundary conditions. In this expression,
 $\gamma(T)$ is the  
vapor-liquid interfacial tension. For $\rho$ near
$\rho_d(T)$, the system is in a state of two-phase coexistence, a
slab-like liquid domain is separated from the vapor via those
interfaces. 
Coexisting gas and liquid phases
have the same free energy. Therefore, $\Delta F$ is the
free energy of the interface.
It has been amply verified for a variety of systems
\cite{B-82,BHN-93,MBO-95,HR-95,PP-00,VHB-05} that the relation \cite{B-82}

\begin{equation} \label{eq14}
P_{\mu, T}(n_d) / P_{\mu, T} (n_{\rm coex}) \propto \exp [-2
\gamma (T) L^2 /k_BT]
\end{equation}

(where $n_d=\rho_d(T) L^3$ and $n_{\rm coex}=\rho^{(1,2)}_{\rm
coex} L^3$) is a valid description of the simulation results, and
can be used to extract rather accurate estimates for $\gamma(T)$.

Close to $T_c$ the estimates for $\rho^{(1)}_{\rm
coex}$, \,$\rho^{(2)}_{\rm coex}$, $\rho_d,$ and $\gamma(T)$ 
suffer from systematic finite size effects. It turns out, however,
that the finite size effects for $\rho_d$ are numerically rather
small. Therefore, the critical density $\rho_c$ can be
estimated from $\rho_c=\rho_d (T_c)$ with
Eq.~(\ref{eq13}).  $\rho^{(1)}_{\rm coex}$ and $\rho^{(2)}_{\rm
coex}$  are just the peak values of the density resulting 
from the equal area rule at $T_c$. (We note that $\rho^{(2)}_{\rm coex} (T_c) >
\rho^{(1)}_{\rm coex} (T_c)$ for any finite $L$. The peak values
only merge into a single point $\rho_c$ at T$_c$ in the thermodynamic
limit.)

The behavior of the density near the critical point can then be
obtained, too. In the critical region the critical exponent $\beta$
has to take the value $\beta=0.325$ of the Ising model
universality class \cite{Z-96}

\begin{eqnarray} \label{eq15}
\rho^{(1)}_{\rm coex} - \rho_d(T) &=& -\hat{B} (1-T/T_c)^\beta \, \nonumber\\
 \rho^{(2)}_{\rm coex} - \rho_d(T) 
 &=& + \hat{B}(1-T/T_c)^\beta  .
 \end{eqnarray}

Here, the critical amplitude $\hat{B}$ can be estimated by fitting
the actual simulation data in the range $0.02 \leq1-T/T_c \leq
0.1$ to Eq.~(\ref{eq15}). Note that the left boundary of this
interval is chosen such that for the typical linear dimensions,
 finite size effects on the peak position estimates for
$\rho^{(1)}_{\rm coex}$, $\rho^{(2)}_{\rm coex}$
  are still very small. The right boundary of
the interval is chosen in order to justify the neglect of
correction terms to the leading term written in Eq.~(\ref{eq15})
which only describes the asymptotic behavior in the
limit \cite{Z-96} $1-T/T_c \rightarrow 0$.

Our data for the
coexistence curve and interfacial tension were derived from an
elongated box $L\times L\times 2L$ with
size $L= 9 \sigma$ and $L= 6.74 \sigma$ (the latter only very far from the
critical point). The critical points (Figs.\ \ref{fig1}, \ref{fig2})
were computed using cubic boxes
of size $9 \,\sigma$ and $11.3 \,\sigma$. In a few cases, a larger box
$L=13.5 \,\sigma$ was implemented to check the finite size effects.
After coexistence densities were determined, simulations at
coexistence gas density were carried out in the NVT
ensemble to obtain the coexistence pressure from the standard virial
expression.

\section{Numerical Results for Carbon Dioxide: Comparison with Experiment and 
Simulations of  Atomistic Models}

Figs.~\ref{fig3}-\ref{fig5} present the coexistence curve, the
vapor pressure at coexistence, and the interfacial tension as a
function of temperature, and compare them to pertinent
experimental data. \cite{NIST} 
If quadrupolar interactions are neglected ($q_c=0$), a distinct
discrepancy between the experimental data and the simulations can
be observed for the liquid branch of the coexistence 
curve. \cite{VMMB-04,BMVM-05} Agreement with experiments improves
considerably for the isotropic quadrupolar model. A value of $q_c=0.387$
was used which corresponds to the experimental value of
the CO$_2$ quadrupolar moment $|Q|=4.3$ D$\mathrm{ \AA}$
(Eq.~(\ref{eq10})) as discussed above.
It is also very
gratifying that both coexistence pressure (Fig.~\ref{fig4}) and
interfacial tension (Fig.~\ref{fig5}) are in almost perfect
quantitative agreement with experimental data, although for these
quantities there is no adjustable parameter available whatsoever.
In particular, the interfacial tension for $q_c=0$ deviates from the
experimental data rather distinctly, while for $q_c=0.387$ there
is excellent agreement.

A small but systematic discrepancy  is still present
for the liquid branch of the coexistence curve
(Fig.~\ref{fig3}). 
 Hence, we have also tried to take $q_c$ as an
adjustable parameter  to optimize the agreement
between the simulated coexistence curve and the experiments.
The rationale for doing so is twofold: first, there is a
considerable uncertainty in the experimental value for $Q$,
leading to a 30\% uncertainty in $q_c$ - it is not even clear that
the value of $Q$ for CO$_2$ in the vapor phase and in the liquid
are exactly the same. Secondly, it might be better
to choose an {\it effective value} for $Q$
because our spherically symmetric model (Eq.~(\ref{eq5}))
 is a rather incomplete description for the interactions 
between  elongated CO$_2$ molecules. 
In principle, the systematic coarse-graining of a chemically 
realistic model could lead to some effective value for $Q$,
which is larger than the experimental one.

Thus, Figs.~\ref{fig3}-\ref{fig5} also include  some simulation
results for a second choice of $q_c$, namely $q_c=0.470$.
Fig.~\ref{fig3} shows that now the agreement between simulation
and experiment for the liquid branch of the coexistence curve is
 better than for $q_c=0.387$, but for the vapor branch
it is slightly worse. The same slight deterioration of the
agreement can also be observed for the coexistence pressure
(Fig.~\ref{fig4}) and the interface tension (Fig.~\ref{fig5}). 
We conclude that  an absolutely
perfect agreement between any  simplified model, such as
 Eq.~(\ref{eq5}), and a real system simply cannot be expected. 
Some uncertainty about the optimum choice of the
parameters of such a coarse-grained model is simply inevitable.
Actually, the level of agreement between experiment
and our model is very good for both choices of $q_c$. 
This is gratifying, since the model will serve as
an excellent starting point for the coarse-grained
modeling of various polymer solutions containing CO$_2$ as a
solvent.

A model of the type of
Eq.~(\ref{eq5})  (named isotropic multipolar or IMP) was also used in 
Ref.\ \onlinecite{MG-03,AM-03} 
and the vapor-liquid
coexistence curve of CO$_2$ was determined 
with  temperature quench MD
techniques. \cite{GM-02} 
The simulation results of Ref.\ \onlinecite{AM-03} are  reported in 
Fig.\ \ref{fig3} (see $\circ$), too.
Although  large systems
were used,  error bars in the
determination of the coexisting densities using NEMD are large in comparison
with ours as discussed above. (Errors for our simulations are smaller than
the size of the symbols and therefore not shown in Figs.~\ref{fig3}-~\ref{fig5}.)
 We also note that Ref.\ \onlinecite{AM-03} uses Lennard-Jones
parameters that differ significantly from ours, namely
$\varepsilon/k_B=215.0$ K and $\sigma$=3.748~\AA\, while we use
$\varepsilon/k_B=252.8$  and
$\sigma$=3.785~\AA\,
for  $|Q|=4.3$ D$\mathrm{ \AA}$.
This is mainly related to the
larger cutoff radius of $4\, \sigma$ used in Ref.\ \onlinecite{AM-03}, which
increases the critical temperature.
Our agreement with experimental results  
(i.e., coexistence curve Fig.\ \ref{fig3},
coexistence pressure Fig.\ \ref{fig4} and isobar Fig.\ \ref{fig7.5})
is, however, clearly very good because our grandcanonical
simulations allows for a very precise determination of the critical point.

Let us ask how our simulation results for the coarse-grained model
compare to the results obtained for  atomistic models 
of CO$_2$. Figs.~\ref{fig6}, \ref{fig7} and \ref{fig7.5} present 
such comparisons
for the coexistence densities and pressures with some results available
in literature.
The EPM  model \cite{HY-95} (denoted by + in Figs.\ 
\ref{fig6} and \ref{fig7})
overestimates  the vapor density at coexistence and underestimates the
coexistence pressure systematically, while the liquid densities are
underestimated only for $T \leq 260$K. For $T
\geq 280$K, the liquid densities of the atomistic simulation are
too large due to the overestimation of $T_c$. When the atomistic
model is rescaled (EPM2) \cite{HY-95} so that the critical temperature and density are         
matched (denoted by $\lhd$ in Figs.\ \ref{fig6} and \ref{fig7}), the agreement 
between the model calculation and
experiment is almost as good as for our coarse-grained model.
However, the rescaled data for the coexistence pressure  are
slightly but systematically too large. The coexistence line for the EPM2 model
has also been obtained  in Ref.\ \onlinecite{VHRPM-00}, in agreement with the previous
work. \cite{HY-95} In Fig.\ \ref{fig7.5} we include
simulation results of Ref.\ \onlinecite{VHRPM-00} for the EPM2 model for the supercritical 
isobar (200 bar). 
The both models work very good in the supercritical region, although
the coarse grained model gives
slightly better
agreement with experimental data for both choices of $q_c$  used in 
this work. Recently,
\cite{ZD-05} another optimized version of the EPM2 model
has been proposed in which
the atomistic energies, lengths and charges  
have been rescaled to optimize  agreement
with the coexistence experiments.
As a consequence, the agreement with experimental results is very good, 
in particular for the coexistence pressure (see $\Diamond$ in Fig.\ 
\ref{fig7}). Simulations  fit  the experimental curve perfectly below
270K, while for higher temperature small deviations appear.
In Ref.\ \onlinecite{SVHF-01}, two center Lennard-Jones models which include a quadrupolar 
point have been studied extensively, and coexistence densities and pressure
were obtained. Tuning  atomistic
parameters,  the agreement with the experimental curve 
has been optimized\cite{VSH-01} without any physical input. As a result,
a quadrupolar moment for CO$_2$ predicted in Ref.\ \onlinecite{VSH-01} equals 
$|Q|$=3.7938 D$\mathrm{ \AA}$    which is quite off from the experimental value 4.3
D$\mathrm{ \AA}$. Finally, there is also a recent simulation,
 \cite{BHS-07} which uses two ab-initio potentials named
BBV\cite{BBV-00} (denoted by $\Box$ in Figs.\ \ref{fig6} and \ref{fig7}) and
SAPT-s\cite{BSJJSKWR-99} (denoted by $\circ$ in Figs.\  \ref{fig6} and \ref{fig7}).
Results are quite off  the respective experimental values, but 
unlike to the previously mentioned models,  no fitting procedures have been 
applied.
No data on the interfacial
free energy of the atomistic model are available so far to which
we could compare our results.
Figs.~\ref{fig6} and \ref{fig7} demonstrate that the rescaled atomistic
model agrees better with experiment than the
simple LJ model which ignores the quadrupolar interaction
completely. \cite{VMMB-04,BMVM-05} However, in comparison with the present
model (Eq.~(\ref{eq5})), the atomistic models offer no advantages,
even if one rescales the parameters to match the critical point.
In fact, the use of Coulomb interactions in the atomistic models
makes the code considerably slower.

\section{Other quadrupolar fluids}

For a detailed discussion on how to derive simulation parameters
for an arbitrary quadrupolar substance, the reader is referred to appendix 
\ref{appB}. Here we would like to focus on testing the model
for other quadrupolar substances.
Using literature data for $Q$, $T_{c,\mathrm{exp}}$ and 
$\rho_{c,\mathrm{exp}}$ for various
molecular fluids, we can  use our master curves
(Fig.~\ref{fig1}) to predict the value of $q_c$
and describe these fluids with our model,
Eq.~(\ref{eq5}). 
Inserting Eq.\ (\ref{eq8}) into Eq.\ (\ref{eq6})
we obtain
\begin{eqnarray}
q_c &=& {Q^4 \over (k_\mathrm{B} T_\mathrm{c,exp})^2}
\Bigg[{\rho_\mathrm{c,exp} N_\mathrm{A} \over M_\mathrm{Mol}}
\Bigg]^{10/3} {T^*_c(q_c)\over \rho^*_c(q_c)^{10/3}}
\nonumber\\
&\equiv& \lambda_\mathrm{exp}  {T^*_c(q_c)\over \rho^*_c(q_c)^{10/3}} \quad .
\label{eqqB1}
\end{eqnarray}
Note that  $\lambda_\mathrm{exp}$ contains all the experimental parameters
which are required to define the model.  Fig.\ \ref{fig8} plots $q_c$
as a function of  $\lambda_\mathrm{exp}$ for CS$_2$, N$_2$, CO$_2$, C$_2$H$_2$,
and C$_6$H$_6$.

 One  recognizes immediately that for N$_2$ and
CS$_2$ the effects of the quadrupolar interactions can only be
minor, since $q_c$ is very small. Consequently, the simple LJ
model (where quadrupolar effects are completely neglected) should
be a reasonable description of the coexistence densities,
coexistence pressures, and interfacial free energies of those
fluids. Fixing the LJ parameters for N$_2$ via $T_c$ and $\rho_c$ as done in
our previous work, \cite{VMMB-04,BMVM-05} we can  test immediately this
hypothesis  (Fig.~\ref{fig9}).  As expected, the
deviations from the simple LJ fluid are indeed much less
pronounced than for CO$_2$. Note  that these deviations
between the measured and the predicted coexistence curves for these
fluids with small $q_c$ are comparable to the deviations found
between the simple Lennard-Jones coexistence curve and the
experimental results for noble gases such as Ne, Ar, Kr and Xe. These systems
 are considered to be the best experimental realization of
a Lennard-Jones fluid (Fig.~\ref{fig12}). In a
rescaled representation ($T/T_c$ plotted vs. $\rho/\rho_c$), however, 
the
various noble gases do not exactly satisfy a ``law of corresponding
states''. This implies that even for  systems with perfectly
spherical atoms, a description in terms of (classical) point
particles interacting with purely pairwise potentials of the same
functional form (with one parameter for the strength and another
for the range) is not strictly valid.

 These small deviations may
be due to the need for three-body forces\cite{PGCBBA-87}, or quantum
corrections which account for differences in atomic masses.
The inclusion of the three body interaction is computationally 
extremely expensive. Indeed in the evaluation of the total energy 
of the system one would need to evaluate a total number of contributions
that scales like N$^3$ instead of  N$^2$ as for the two body 
interactions (N, being the total number of molecules). For this reason
the inclusion of such effects in our simple (and cheap) modeling
is out of discussion, especially in view of more complicated 
polymer solution applications. There are several attempts 
\cite{RS-03,WS-06} which try to capture the three body interaction
in an effective (density dependent) two body interaction. These methods
cannot be used in non homogeneous fluids and generally where strong density
fluctuations  are present, like near the critical point.
The fact that the method proposed in this work is based on a 
careful investigation of the critical points of the coarse grained models
invalidates the  scheme proposed in Ref.\  \onlinecite{RS-03,WS-06}. However
in Ref.\ \onlinecite{PGCBBA-87} a quantitative estimate of the effects
of the three body interaction is given starting from a careful 
scaling investigation of the rectilinear diameter (\ref{eq13})
\begin{equation}
{\rho_d(T)\over \rho_d(T_c)} = 1 + A_{1-\alpha} \Bigg(1- {T\over T_c} \Bigg)^{(1-\alpha)} 
+ A_1  \Bigg(1- {T\over T_c} \Bigg) +\cdots
\label{eq18_v1}
\end{equation} 
with $\alpha\approx 0.11$. The authors shows that  in Eq.\ (\ref{eq18_v1})
$A_{1-\alpha}$ is related to the field mixing effect (indeed the lack 
of the particle hole symmetry), while  $A_1$ could give an estimate
of the three body interaction. A Mean Field van der 
Waals equation predicts\cite{PGCBBA-87} $A_1 = 2/5$. 
Deviations
of the experimental data from this  law of corresponding states ($A_1 = 2/5$)
 are supposed to be related to the emergence of another energy
scale like that of  three body interactions. Fig.\ 4 of Ref.\
\onlinecite{PGCBBA-87} suggests (for CO$_2$) $A_1 \approx 0.95$ which
differs significantly from the van der 
Waals value $A_1=0.4$ but is comparable with other fluids in particular
Xenon. Comparing now  the   predictions  
 for Xenon (Fig.\ \ref{fig12}) and Carbon Dioxide (Fig.\ \ref{fig3}),
one can easily conclude that in our case  the quadrupolar
interactions are much more relevant than three body interactions.

For the sake of completeness, in Fig.\ \ref{fig12} we have also included the
full LJ potential.
In an unscaled representation one would of course observe large 
differences between the results for the full Lennard-Jones potential 
and those for the cut-and-shift Lennard-Jones potential. In a scaled 
representation these differences vanish almost 
completely (except for small densities on the gas branch of the binodal) 
so that due to its computational efficiency the cut-and-shift potential 
should be preferred in coarse-grained simulations.

The case of benzene (C$_6$H$_6$) is even more interesting. Depending on
which  experimental value  is adopted for $Q$, one finds $q_c$ in the
range from $q_c=0.121$ (for $Q=10$ D$\mathrm{ \AA}$) to $q_c =0.247$ (for
$Q=12$ D$\mathrm{ \AA}$). Fig.~\ref{fig13} compares experimental
values for the coexistence densities, coexistence pressure and
interfacial tension with our predictions, using 
$q_c=0.247$.
In this case we also observe a clear improvement  of the agreement
with experimental data with respect to the pure Lennard Jones case
($q_c=0$ in Fig.\ \ref{fig13}).  
Deviations  are 
of the same order of magnitude as for nitrogen (Fig.\ \ref{fig9}) 
and noble gases (Fig.\ \ref{fig12}).

\section{Predictions for the equation of state resulting from Perturbation Theory
(PT)}

In this section we present  results for  coexistence
densities  and coexistence pressures
(Fig.~\ref{fig15}), which were obtained analytically using an equation of
state  \cite{MMVB-00} in the Mean
Spherical Approximation (PT-MSA) \cite{MMVB-00}. As is well-known
\cite{HM-86} such approaches should work well at temperatures and
densities away from the critical region. \cite{HM-86} This
expectation is reconfirmed by our results
(Fig.~\ref{fig15}), which show good agreement at
temperatures below 0.9 $T_c$. For 0.9 $T_c \leq T \leq 1.2
\ T_c$, there are distinct deviations between simulations and theory 
because PT-MSA
overestimates the critical temperature by about 10\% and furthermore 
the slope of the binodal in the critical region is mean-field-like in 
PT-MSA and Ising-like in the simulation.
For low temperatures the deviations are quantitatively smaller, 
however, the MC 
results and PT-MSA results cross
at $T^*\approx 1$ (if $q_c=0.387$) and  $T^*\approx 1.05$
(if $q_c=0.470$) on the liquid branch.
Note that
our comparison involves no adjustable parameter whatsoever.  
For many practical applications one will be interested in the
temperatures and/or densities outside the critical region. Hence, the
results shown in Fig.~\ref{fig15} are 
encouraging in that a  relatively simple analytic
method such as PT-MSA (see Appendix A for some details on this
method) works well as a description of the equation of state for
molecular fluids like CO$_2$ away from the critical region 
if an isotropic quadrupolar
interaction is included. To some extent this minimizes
the need for massive Monte Carlo
(MC) efforts to explore phase space.
Even though MC simulations are required to determine $\varepsilon$,
$\sigma$ and $q_c$ from $T_{c,\mathrm{exp}}$, $\rho_{c,\mathrm{exp}}$
and Q, the results are already contained in Fig.\ \ref{fig1}
and Eqs.\ (\ref{eq9a},\ref{eq9b}).
Therefore, no new
efforts with MC simulations will be needed for any future applications of
PT-MSA in the context of our model.

\section{Conclusions}

In the present work, the thermodynamic properties of a coarse-grained model for
quadrupolar fluids were investigated. A particular emphasis was put on the question
to which extent the equation of state and the interfacial tension between
coexisting vapor and liquid phases can be described accurately. 

The aim of this work hence is not a chemically detailed modeling of 
quadrupolar fluids on an atomistic level, but rather to derive a model
which is bot simple and accurate enough that it can serve as a 
starting point for the description of binary fluid mixture, solvents
in polymer solutions, etc.. Obtaining efficient models for such 
purposes is a topic of great current interest.

As experimental input parameters, our description only requires 
knowledge of the experimental critical temperature
$T_{c,\mathrm{exp}}$ and the critical density 
$\rho_{c,\mathrm{exp}}$ of the fluid and the  experimental quadrupole
moment $Q$ of the molecule. 
The quadrupolar interaction is treated in a spherical approximation \cite{SRN-74,MG-03} 
which can be derived from thermodynamic perturbation theory. This leads to an effective
potential proportional to $Q^4$/($Tr_{ij}^{10}$), where $T$ denotes 
the temperature and $r_{ij}$
the distance between the centers of mass of molecules $i$ and $j$. The application of the
isotropic quadrupolar interaction is mainly motivated by the desire to have a very fast
simulation code.
Steric and dispersion forces are simply modeled by a Lennard-Jones
potential involving parameters $\varepsilon$ and $\sigma$, which define the strength and the range
of the interaction, respectively. In practice, the potential is cut and shifted to zero at a
cutoff range $r_c=2\sqrt[6]{2}$, which is again motivated by our desire to speed up
calculations. We also provide evidence that this particular approximation
mostly affects the conversion factor from $\varepsilon$ to experimental
 temperature and hence does not alter results
significantly.

For the description of a real system, simulation parameters $\varepsilon$, $\sigma$ and
$q_c=Q^{4}/(\varepsilon\sigma^{10}k_{B}T_{c,\mathrm{exp}})$ need to be determined from experimental
values $T_{c,\mathrm{exp}}$, $\rho_{c,\mathrm{exp}}$ and $Q$ in physical units. 
To address this problem, we have
determined master curves $T^*_{c}(q_{c})/T^*_{c}(0)$ and $\rho^*_{c}(q_{c})/\rho^*_c(0)$
as a function of $q_c$ (Fig.~\ref{fig1}, Eqs.~\eqref{eq9a},\eqref{eq9b}). This task is performed
easily using grand-canonical Monte Carlo simulations\cite{B-97,LB-05,W-97} in combination with reweighting,
successive umbrella sampling\cite{VM-04} and finite-size scaling methods\cite{W-97,F-71,B-92}. With modest computational
effort, these master curves are determined with a relative accuracy which is distinctly
better than 1$\%$.

Carbon dioxide is a prototype of a linear elongated molecule 
with a rather large quadrupole
moment. Comparing our predictions for the coexistence curve, vapor pressure at coexistence
and interfacial tension with 
corresponding experimental data \cite{NIST}, we found encouragingly  good
agreement (Figs.~\ref{fig3},\ref{fig4},\ref{fig5}). 
Note that after having fixed the scales for temperature
and density via $\varepsilon$ and $\sigma$, no further parameters need to be adjusted, 
neither for
the pressure (Fig.\ref{fig4}), nor for the interfacial tension (Fig.~\ref{fig5}). 
The level of agreement which
we have achieved is clearly nontrivial. However, the inclusion of quadrupolar effects is
essential to the model and agreement with experiments deteriorates significantly if CO$_2$
is described by a Lennard-Jones particle without quadrupole moment.

Our model produces  rather accurate off-critical isotherms, too. 
As expected, the comparisons also reveal 
 small discrepancies, since such a simple model cannot 
be absolutely perfect. 
However, a more realistic model, based on an all atom description of CO$_{2}$
which involves considerably more complicated potentials, performs distinctly 
worse in comparison to our model --
 except if experimental critical parameters are used to
 empirically re-calibrate the atomistic potential.
In our view, such a procedure looses the advantage of a fully predictive
 modeling that does not need experimental input.
Complicated atomistic models also
 lead to rather slow simulation programs (partial charges require to deal with
 rather long range coulombic 
interactions, etc.). While such models 
may still  be  manageable for the simulation of pure fluids, 
their drawbacks become clearly apparent when the approach is extended to
 binary or ternary fluids.
In mixtures, a large control parameter space needs to be explored and several
 phase separations may 
compete with each other, leading to very involved phase diagrams.

We emphasize that our successful description of carbon dioxide is by no means
 accidental. As a counterpart, we also consider nitrogen,
 a fluid with a considerably  smaller quadrupole moment.
In this case,  a simple Lennard-Jones model with no quadrupolar forces should
provide an equally good description, and in fact it does.
The deviations are comparable to the deviations found between the coexistence
curve of "Lennard-Jonesium" and those of various noble gases (that do not
superimpose precisely in a re-scaled representation shown in Fig. \ref{fig12} 
either.)
This indicates that a simple pair potential with two parameters for the scales of energy and
 range does not suffice even for these prototypes of simple spherical atoms. 

As a further example, we also present a comparison between our model and
 experimental data for benzene  (C$_6$H$_6$).
Again, the agreement is very good. This result is of great interest, since the
 shape of the benzene molecule differs considerably from CO$_2$, 
consisting of a disk rather than an elongated ellipsoid.

A very interesting question  is the extent to which this concept can 
 actually  be carried over from simple fluids to binary mixtures and
 polymer-solvent systems. Are interactions between different types of
 molecules   captured by simple Lorentz-Berthelot mixing rules, when one
 describes the pure constituents with  the quality of the present work? We
 shall address this very interesting and potentially practically 
useful question in a forthcoming paper. 
We also hope that the present work will stimulate some analytical research, 
starting from general statistical mechanics of fluids, to provide a better 
theoretical understanding for the high accuracy of our approach. We also 
point out that the knowledge of the appropriate parameters $\varepsilon$, 
$\sigma$ and $q_{\rm c}$ allows a rather accurate description of the equation 
of state by liquid-state perturbation theories at state points sufficiently 
away from the critical region (Sec. V).
\\
\\
{\bf ACKNOWLEDGEMENTS}\\
CPU times was provided by the NIC J\"ulich and the  ZDV Mainz.
We would like to thank F.Heilmann and H.Weiss of
BASF AG (Ludwigshafen) for fruitful discussions and E.M\"uller (London) for 
CO$_2$ simulations data of IMP model.
BMM would also like to acknowledge BASF AG (Ludwigshafen) 
for financial 
support and J. Horbach for useful discussions. 
LGM wishes to acknowledge support from Ministerio de Educacion y
Ciencia 
(project FIS2007-66079-C02-00) and Comunidad Autonoma de Madrid
(project MOSSNOHO-S0505/ESP/0299).

\appendix
\section{Mean Spherical Approximation (MSA) predictions}\label{appMSA}
In this appendix we want to give some technical details concerning
the analytical predictions presented in this paper. 
For more details we refer to the original literature. In particular, 
the equation of state (EOS) used in this work is a straightforward 
generalization of the EOS given in appendix B of Ref.\ \onlinecite{MMVB-00} for
the case in which four Yukawa tails are used instead of two.
We follow the strategies of  Refs.\ \onlinecite{TL-93,TTL-97} in which 
the Ornstein-Zernike (OZ) equation is solved in a first order MSA closure.
The general idea \cite{HM-86} is to divide the potential
into a repulsive part (that becomes the reference potential) plus
 a perturbative attractive part
\begin{eqnarray}
U_\lambda(r)= \left\{ \begin{array}{ll} 
  U_\mathrm{rep}(r) & \textrm{if $r<\sigma_0$}\\
  \lambda U_\mathrm{att}(r) & \textrm{if $\sigma_0<r<r_\mathrm{cut}$},
\end{array} \right.
\label{att-rep}
\end{eqnarray}
where $U(\sigma_0)=0$, $ U_\mathrm{rep}(r)>0$, $ U_\mathrm{att}(r)<0$ and 
$\lambda$ is the  perturbative parameter. 
The reference system ($\lambda = 0$)
is modeled by hard spheres
 with a proper radius $d_\mathrm{HS}$,  \cite{BA-67} computed
using $U_\mathrm{rep}$ \cite{BA-67}.
In order to get corrections to the reference free energy $A_\mathrm{ref}$, 
a systematic expansion in  $\lambda$
is developed (the general expression for $A-A_\mathrm{ref}$ is
  standard
and can be found for example in Ref.\ \onlinecite{MMVB-00} (Eq.\ B5)).
 The explicit solution up to second order in $\lambda$
 has been obtained in. Refs.\ \onlinecite{TL-93, TTL-97}
The key point developed in Ref.\ \onlinecite{TTL-97} is
to fit $U_\mathrm{att}$ with a couple of Yukawa tails. In the case
of the LJ potential this yields
\begin{eqnarray}
U^\mathrm{LJ}_\mathrm{att} &\approx& -c_1{e^{-z_1(r -\sigma_0)}\over r}
+c_2 {e^{-z_2(r -\sigma_0)}\over r} 
\nonumber\\
&\equiv & {\cal Y}^\mathrm{LJ}(c_i,z_i,\sigma_0;r).
\label{vlj_yu}
\end{eqnarray} 
 In this work the LJ part of the potential
is fitted using the same Yukawa tail as reported in Ref.\ \onlinecite{MMVB-00} (Eq.\  B6).
Equation (\ref{vlj_yu})
 allows us to invert some Laplace transforms that are present in the 
Tang-Lu solution \cite{TL-93} and to obtain an analytical expression for the
free  energy which is explicitly
given in Eq.\ B7-B10 of Ref.\ \onlinecite{MMVB-00} for the apolar-fluid case $q=0$.

For the general case $q\neq 0$, Eq.\ (\ref{att-rep}) will 
induce the same decomposition on both the LJ part and
quadrupolar part of the potential 
\begin{eqnarray}
U_\mathrm{att(rep)} &=& U^\mathrm{LJ}_\mathrm{att(rep)}-{7\over 20}
q  \,  U^\mathrm{IQQ}_\mathrm{att(rep)}.
\label{splitVstart}
\end{eqnarray}
In (\ref{splitVstart}) 
we have used two more Yukawa tails to fit the 
quadrupolar interaction $U^\mathrm{IQQ}_\mathrm{att}$ 
\begin{eqnarray}
U^\mathrm{IQQ}_\mathrm{att} &\approx& -c_3{e^{-z_3(r -\sigma_0)}\over r}
+c_4 {e^{-z_4(r -\sigma_0)}\over r} 
\nonumber\\
&\equiv& {\cal Y}^\mathrm{QQ}(c_i,z_i,\sigma_0;r).
\label{vqq_yu}
\end{eqnarray} 
Because $q$ ($=q_c T_c/T$) is factored out in
(\ref{splitVstart}), $c_{3,4}$ and $z_{3,4}$ do not depend
on temperature $T$.
This is an important simplification because  using   
(\ref{vlj_yu}) and (\ref{vqq_yu})
we can get an immediate fit for $U_\mathrm{att}$ (\ref{splitVstart})
for every $q$ and $T$
\begin{eqnarray}
U_\mathrm{att} \approx {\cal Y}^\mathrm{LJ}(c_i,z_i,\sigma_0;r) -{7\over 20}
q\, {\cal Y}^\mathrm{QQ}(c_i,z_i,\sigma_0;r).
\label{lastfit}
\end{eqnarray}
By using the previous fit (\ref{lastfit}) and extending
 Eq.\ B7-B10 in Ref.\ \onlinecite{MMVB-00}  to the case in
which more than two Yukawa expressions are used to fit the potential,
we have obtained the desired EOS used in the present work.

\section{Determination of simulation parameters}\label{appB}

Simulation parameters $\varepsilon$, $\sigma$ and q$_c$ are needed to convert simulation
units into experimental units. Knowledge of q$_c$, or rather $q=q_c \cdot T_{c} / T$ is also required 
as input before a simulation can be started.
In Table \ref{table1}, we have collected the simulation parameters for the quadrupolar substances 
mentioned in the paper. However, we would also like to convey some hands-on 
knowledge on how to calculate these parameters and extend the 
model to substances not listed in Table \ref{table1}. Furthermore, we provide fitting curves (Table \ref{table2}) which allow us to
determine the phase diagram of an arbitrary substance without additional MC simulations.

For q$_{c}$=0, $\varepsilon$ and $\sigma$ can be determined directly from the 
critical temperature T$_{c}$ and the critical density $\rho_{c}$ using Eq. \eqref{eq8}.
For $q_{c}\ne0$, the location of the critical point itself depends on $q_c$.
Therefore, $\varepsilon$ and $\sigma$ also depend on $q_c$ (Eq. \eqref{eq8}), 
and 
a simple iteration procedure can be formulated.
Starting with $q_{c}$=0, $T_{c}$ and $\rho_{c}$ are computed using the master curves from
Eqs. \eqref{eq9a} and \eqref{eq9b}. 
From these results,  $\varepsilon$ and $\sigma$ are
determined with Eq. \eqref{eq8} and a new value for $q_{c}$ with Eq. \eqref{eq6}.
The iteration is repeated until $q_c$, $\varepsilon$ and $\sigma$ converge. 
Usually,
around 5-10 iterations are sufficient to obtain simulation parameters with 
good accuracy without any additional simulations. In the following, we present a pseudo-code
for our CO$_2$ calculations which can be extended to any quadrupolar substance by substituting
experimental values for Q=4.3~$D\mathrm{ \AA}$, T$_c$=304.1282~$K$, and $\rho_c$=10.6249~$mol/l$: 
\\
\\
\noindent
\underline{Initialize variables}
\\
\noindent
Q = 4.3  \hspace{3.45cm} /* \hspace{0.1cm} $D\mathrm{ \AA}$ \hspace{0.1cm}*/ \\
Q = Q*3.33564*10$^{-40}$;  \hspace{0.8cm} /* \hspace{0.1cm} convert Q to SI units \hspace{0.1cm}*/ \\
T$_\mathrm{c,exp}$ = 304.1282; \hspace{1.6cm} /* \hspace{0.1cm} K \hspace{0.1cm}*/ \\
rho$_\mathrm{c,exp}$ = 10.6249;  \hspace{1.5cm} /* \hspace{0.1cm} mol/l \hspace{0.1cm}*/ \\
T$_\mathrm{LJ}$(q=0) = 0.99821  \hspace{1.15cm} /*  \hspace{0.1cm} critical temperature of simulation for q=0 \hspace{0.1cm}*/ \\
rho$_\mathrm{LJ}$(q=0) = 0.32276 \hspace{0.85cm} /*  \hspace{0.1cm} critical density of simulation for q=0 \hspace{0.1cm}*/ \\
q        = 0;
\\
\\
\noindent
\underline{Iteration}
\\
for (i=0;i$<$20;i++) $\{$\\
\hglue1cm          T        = T$_\mathrm{LJ}$(q=0) * (1 + 0.46111 * q + 0.17571 * q$^2$);  \hspace{0.7cm} /*\hspace{0.1cm} Eq.\eqref{eq9a} \hspace{0.1cm}*/\\
\hglue1cm          density  = rho$_\mathrm{LJ}$(q=0) * (1 + 0.19298 * q); \hspace {2.2cm} /*\hspace{0.2cm} Eq.\eqref{eq9b} \hspace{0.1cm}*/ \\ 
\hglue1cm          epsilon  = T$_\mathrm{c,exp}$ * 1.38065 * 10$^{-23}$ / T; \hspace {4.0cm} /*\hspace{0.1cm} Eq.\eqref{eq8} \hspace{0.1cm}*/ \\
\hglue1cm          sigma    = (rho$_\mathrm{c,exp}$ * 1000 * 6.02214 * 10$^{23}$ / density ) $^{-1/3}$ ; \hspace{0.4cm} /*\hspace{0.1cm} Eq.\eqref{eq8} \hspace{0.1cm}*/ \\
\hglue1cm          Q$_1$       = Q/(sqrt(epsilon*sigma$^5$)); \hspace{2.8cm} /*\hspace{0.1cm} Eq.\eqref{eq6} \hspace{0.1cm}*/\\
\hglue1cm          q        = Q$_1^4$ / ($T*1.237990147*10^{-20}$); \hspace{0.5cm} /* \hspace{0.1cm} T$_\mathrm{sim}$=k$_\mathrm{B}$T$_\mathrm{exp}$/$\varepsilon$, \hspace{0.3cm} (4$\pi\varepsilon_{0}$)$^{2}$ - SI units\hspace{0.1cm} */ \\
\hglue1cm          print T, epsilon, sigma, q;\\
$\}$\\

Alternatively, q$_c$ can also be determined from the fitting curve in Fig.~\ref{fig9}. 
$\lambda_\mathrm{exp}$ is a dimensionless parameter, which already contains all the experimental 
information required to define the model. If all constants are included, $\lambda_\mathrm{exp}$ reduces
to
\begin{eqnarray}
\lambda_\mathrm{exp} = 96.754\cdot 10^{-5}  \frac{Q^{4}}{T_\mathrm{c,exp}^{2}}
( \rho_\mathrm{c,exp} )^{\frac{10}{3}}. &
\label{eqB1}
\end{eqnarray} 
In this equation, one simply needs to plug in experimental values for quadrupolar moment $Q$ in D$\mathrm{ \AA}$, critical temperature 
T$_\mathrm{c,exp}$ in K, and critical molar density $\rho_\mathrm{c,exp}$ in mol/cm$^{3}$. q$_{c}$ can be read off from Fig.~\ref{fig9} or determined 
via the following fit to the curve:  

\begin{eqnarray}
q_c =  \lambda_\mathrm{exp} ( 43.1018 -266.251\lambda_\mathrm{exp}+ 5047.01 \lambda_\mathrm{exp}^2 )
& \qquad & \lambda_\mathrm{exp} \le 0.02
\label{eqB2}
\end{eqnarray}

\noindent
T$_c$, $\rho_c$, $\varepsilon$ and $\sigma$ follow from Eqs.~\eqref{eq9a}, \eqref{eq9b}, and \eqref{eq8}.

\noindent

Finally, we demonstrate how our accumulated simulation data can be used to provide a rough 
estimate of the phase diagram for an arbitrary quadrupolar substance without any 
additional MC simulations. We simulated several values for $q_c$ in the range of 
$0.1 \le q_c \le 0.47$. Four temperatures were considered such 
that  $T^*_i(q_c)/T^*_c(q_c)$ ($i=1, \cdots ,4$) is independent of $q_c$: 
$T^*_1=0.974499\cdot T^*_c$, 
$T^*_2=0.932125\cdot T^*_c$, $T^*_3=0.864337\cdot T^*_c$, 
and $T^*_4=0.813494\cdot T^*_c$. As indicated before, 
critical quantities scale almost linearly with q$_c$ (Fig.\ref{fig1}, Eqs.\eqref{eq9a} and \eqref{eq9b}). During our 
investigations, we observed that this approximation also holds away 
from criticality. The corresponding fitting curves are listed in Table~\ref{table2}.

First, one needs to determine $\varepsilon$, $\sigma$ and q$_c$ for the
substance in question as demonstrated in the previous section.
Vapor and liquid coexistence densities, interface tension and pressure 
at the selected temperature can be computed by inserting q$_c$ 
into the respective fitting curves. The following equations can be used to 
convert the results from simulation units to experimental units:
\begin{equation}
T_\mathrm{exp}= {\varepsilon(q_c) \over k_B} T^*_i,  \quad
\rho_\mathrm{exp,l,g} = \rho^*_{l,g} \frac{M_\mathrm{mol}} 
{N_A \sigma(q_c)^3},\quad
\gamma_\mathrm{exp} = \gamma^*  
\frac{\varepsilon(q_c)}{\sigma(q_c)^2},\quad
p_\mathrm{exp} = p^* \frac{\varepsilon(q_c)}{\sigma(q_c)^3}. 
\end{equation}

\bibliographystyle{unsrt}

\newpage
\clearpage

\begin{table}
\caption{Experimental data and simulation parameters for several 
quadrupolar substances as obtained in the present work}
\label{table1}
\end{table}

\begin{table}
\caption{ Fitting curves to determine coexistence properties for an arbitrary
  quadrupolar substance at selected temperatures $T^*_i(q_c)/T^*_c(q_c)$ ($i=1, \cdots ,4$):         $T^*_1=0.974499\cdot T^*_c$,
$T^*_2=0.932125\cdot T^*_c$, $T^*_3=0.864337\cdot T^*_c$,                         and $T^*_4=0.813494\cdot T^*_c$ (see text) }
\label{table2}
\end{table}

\begin{figure}
\caption{Master curves: normalized critical temperature 
$T_c^*(q_c)/T_c^*(0)$, 
normalized critical density 
$\rho^*_c(q_c)/\rho^*_c(0)$,  
and normalized critical  pressure
$p_c^*(q_c) / p_c^*(0)$ 
plotted versus the quadrupolar parameter $q_c$. 
Symbols represent simulation data, curves are
the interpolating functions (Eqs.\ (\ref{eq9a}) and (\ref{eq9b})) and 
$p_c^*(q_c) / p_c^*(0) = (1+ 0.67423 \, q_c + 0.274349 \, q_c^2)$ 
with $p_c^*(0)=0.087221$.}
\label{fig1}       
\end{figure}

\begin{figure}
\caption{Second and fourth order cumulants $U_2$, $U_4$ plotted for $q=0.3$ versus
$T^*=k_BT/\varepsilon$ for three choices of $L$. Broken horizontal
values indicate the theoretical values established for the Ising
universality class. \cite{LB-05,W-97} 
From the intersections one can conclude $T^*_c=1.152\pm 0.003$
for this particular case.
Inset: the slope of the fourth order
cumulants ($Y_1$) as a function of the box size,
on a log-log scale. The data points fall on a straight line with a
slope
 equal to 1.584 in agreement with the finite size prediction
$1/\nu$, with $\nu\approx 0.630$ for the Ising universality class.
\cite{Z-96}}
\label{fig2}       
\end{figure}

\begin{figure}
\caption{Coexistence curve of CO$_2$ plotted in the
temperature-density plane. The broken curve denotes the experimental data
(from NIST \cite{NIST}), the full curve is the result for the LJ model 
without quadrupolar
interactions\cite{VMMB-04}. 
Solid square denotes the critical point of CO$_2$.
($\times$) and ($\ast$) are the
results of the present $\mu VT$ work for two choices of $q_c=q(T_c)$
as indicated in the figure. ($\circ$) are the results of the 
spherical averaged model investigated in Ref.\ \onlinecite{AM-03}. }
\label{fig3}       
\end{figure}

\begin{figure}
\caption{Coexistence pressure of CO$_2$ plotted vs. temperature.
The broken curve denotes the experimental data, \cite{NIST} the full 
curve:
the results for the LJ model without quadrupolar interactions.
($\times$) and ($\ast$) are the results of the present NVT work for two
choices of $q_c=q(T_c)$ as indicated in the figure.}
\label{fig4}       
\end{figure}

\begin{figure}
\caption{Interface tension $\gamma(T)$ of CO$_2$ plotted vs.\
temperature. The broken curve denotes the experimental data (from NIST
\cite{NIST}), the full curve: the results for the LJ model without
 quadrupolar
interactions.\cite{VMMB-04} ($\times$) and ($\ast$) are
the results of the present work for two choices of $q_c=q(T_c)$  as
indicated in the figure.}
\label{fig5}       
\end{figure}

\begin{figure}
\caption{Coexistence curve of CO$_2$ plotted in the
temperature-density plane. The broken curve denotes the experimental data
(from NIST \cite{NIST}), the full curve: the results 
for LJ model without  quadrupolar
interactions\cite{VMMB-04}. 
($\bullet$) denotes the critical point of CO$_2$. 
($\ast$) and ($\times$) denote 
the results of this work for $q_c=0.387$ and $q_c=0.470$,
respectively. ($+$) are the results of the EPM model introduced in 
Ref.
\ \onlinecite{HY-95}. 
($\bigtriangledown$) are the results from Ref.\ \onlinecite{HY-95} for the EPM model with 
flexible molecules, which
give essentially the same thermodynamic properties as the rigid molecules. 
($\lhd$) are the results of Ref.\ \onlinecite{HY-95}
 for the rescaled EPM model
(EPM2). ($\circ$) and ($\Box$)
correspond to simulations \cite{BHS-07} of two ab initio potentials.
\cite{BBV-00,BSJJSKWR-99}}
\label{fig6}       
\end{figure}

\begin{figure}
\vspace*{0.5cm}
\caption{Coexistence pressure of CO$_2$ plotted vs. temperature.
Labeling of curves and symbols is the same as in Fig.~\ref{fig6}.
We also show simulations of an optimized EPM2 model \cite{ZD-05} 
(see $\Diamond$) which is in good agreement with experiments.
We stress  that the nice agreement of our model 
with experiments 
 near the critical 
point is not given a priory because our method  only fixes  the critical
temperature and the critical density.}
\label{fig7}       
\end{figure}

\begin{figure}
\caption{Supercritical isobar for p=200 bar. 
The broken curve denotes  the experimental data. \cite{NIST} 
($\times$) and ($\ast$) are the results of 
the present NVT work for two
choices of $q_c=q(T_c)$ as indicated in the figure.  
($\lhd$) are the
prediction of the atomistic EPM2 model given in Ref.\ \onlinecite{VHRPM-00}.
The coexistence curve near the critical point is also reported.  }
\label{fig7.5}       
\end{figure}

\begin{figure}
\caption{
Estimates for the quadrupolar parameter $q_c$ for various quadrupolar fluids
characterized by parameter $\lambda_\mathrm{exp}$ (Eq.~\eqref{eqqB1}). The corresponding experimentally measured
quadrupole moments $Q$ of these systems are quoted
in brackets (see also Table \ref{table1}).}
\label{fig8}       
\end{figure}

\begin{figure}
\caption{Lennard-Jones results ($q_c=0$) for  N$_2$. From top to bottom: 
coexistence curve in the temperature-density plane, 
vapor pressure vs.\ temperature and interface tension vs.\ 
temperature. Symbols correspond to simulations of a simple
Lennard-Jones model without
quadrupolar moment obtained from 
 $\mu$VT simulations \cite{VMMB-04} (coexistence densities
and interface tensions) and  NVT simulations 
(pressure). The
broken curves denote the experimental data
(from NIST \cite{NIST}).
}
\label{fig9}       
\end{figure}

\begin{figure}
\caption{Coexistence curves ($T/T_c$ plotted vs. $\rho/
\rho_c$) for various noble gases in comparison with 
 the prediction of the cut-and-shifted Lennard--Jones model (LJ)\cite{VMMB-04} 
and the full Lennard-Jones model.\cite{PP-00}}
\label{fig12}       
\end{figure}

\begin{figure}
\caption{New predictions for benzene (C$_6$H$_6$). From top to bottom: 
coexistence curve in the temperature-density plane, 
vapor pressure vs.\ temperature and interface tension vs.\ 
temperature.
The broken curves denote the experimental data,
\cite{NIST} the full curve is the result of the simple Lennard-Jones
model. ($\rhd$) denote the present results which include  an
isotropic quadrupolar interaction  for  $q_c=q(T_c)$ 
corresponding to  $Q=12$ D$\mathrm{ \AA}$.}
\label{fig13}       
\end{figure}

\begin{figure}
\caption{Coexistence densities and coexistence vapor pressure:
a comparison between the MC simulations
and the PT-MSA prediction. 
The two choices of $q_c=q(T_c)$ used in this
work are included 
as indicated.}
\label{fig15}       
\end{figure}

\newpage
\clearpage

\begin{table}
\begin{tabular}{|c|c|c|c|c|c|c|c|}
\hline
Subst.  &  Q [D$\mathrm{ \AA}$] &   $T_{c,\mathrm{exp}}$ [K] & $\rho_{c,\mathrm{exp}}$ [mol/l] & $\lambda_\mathrm{exp}$  & $q_c$  &  $\varepsilon/k_\mathrm{B}$ [K]  &
$\sigma$ [\AA]   \\
\hline
CO$_2$ & 4.3 &  304.1282 & 10.6249 & 0.009430   & 0.387  & 252.829      &     3.785     \\
\hline
CS$_2$ & 3.6 & 552 & 5.78  & 0.0001848  & 0.0080  &   550.95     &     4.528    \\
\hline
N$_2$ & 1.47  & 126.2 & 11.18 & 0.0008864 & 0.038  &    124.208       &  3.642      \\
      &   0   & 126.2  &11.18 &  0    & 0  &  126.426        &  3.633    \\
\hline
C$_2$H$_2$ & 5.5 & 308.3 & 8.913 & 0.013775     &  0.553  &   235.942     &  4.052    \\
\hline
C$_6$H$_6$ & 12  & 562 & 3.9 & 0.0059311    &   0.247 & 500.468    &  5.242   \\
\hline
\end{tabular}
\end{table}
TABLE I

\newpage
\clearpage

\begin{table}
\begin{tabular}{|c|c|}
\hline
Observable  &  Fitting Formula      \\
\hline
$\rho^*_{1,g}$ &  $\approx 0.162$  \\
$\rho^*_{2,g}$ &  $0.099506 - 0.0094827 \, q_c $ \\
$\rho^*_{3,g}$ &  $0.055372 - 0.017106 \, q_c$  \\
$\rho^*_{4,g}$ &  $0.036003 - 0.018  \, q_c$ \\
\hline
$\rho^*_{1,l}$               & $0.49215 + 0.12426 \, q_c + 0.021146 \, q_c^2$ \\
$\rho^*_{2,l}$               & $0.57055 + 0.15313 \, q_c + 0.025081 \, q_c^2$ \\
$\rho^*_{3,l}$               & $0.64597 + 0.1531 \, q_c + 0.09854 \, q_c^2$ \\
$\rho^*_{4,l}$               & $0.68355 + 0.22094 \,  q_c + 0.042765 \, q_c^2$ \\
\hline
$\gamma^*_1$ & $ 0.020384 + 0.016672 \, q_c + 0.027991 \, q_c^2 $ \\
$\gamma^*_2$ &  $ 0.068376 + 0.072064 \, q_c + 0.082864 \, q_c^2 $ \\
$\gamma^*_3$             &  $ 0.16187 + 0.19493 \, q_c + 0.18704 \, q_c^2 $ \\
$\gamma^*_4$             & $ 0.23945 + 0.29931 \, q_c + 0.30352 \, q_c^2$ \\
\hline
$p^*_1$        &  $ 0.075861 + 0.041526 \, q_c + 0.024072 \, q_c^2$ \\
$p^*_2$        &  $ 0.056804 + 0.026873 \, q_c + 0.011408 \, q_c^2 $\\
$p^*_3$        &  $ 0.035115 + 0.0099939 \, q_c + 0.00067637 \, q_c^2 $ \\
$p^*_4$        &  $0.023617 + 0.0010425 \, q_c - 0.0009939 \, q_c^2 $ \\
\hline
\end{tabular}
\end{table}
TABLE II

\newpage
\clearpage
\begin{figure}
\includegraphics[angle=-90,scale=0.55]{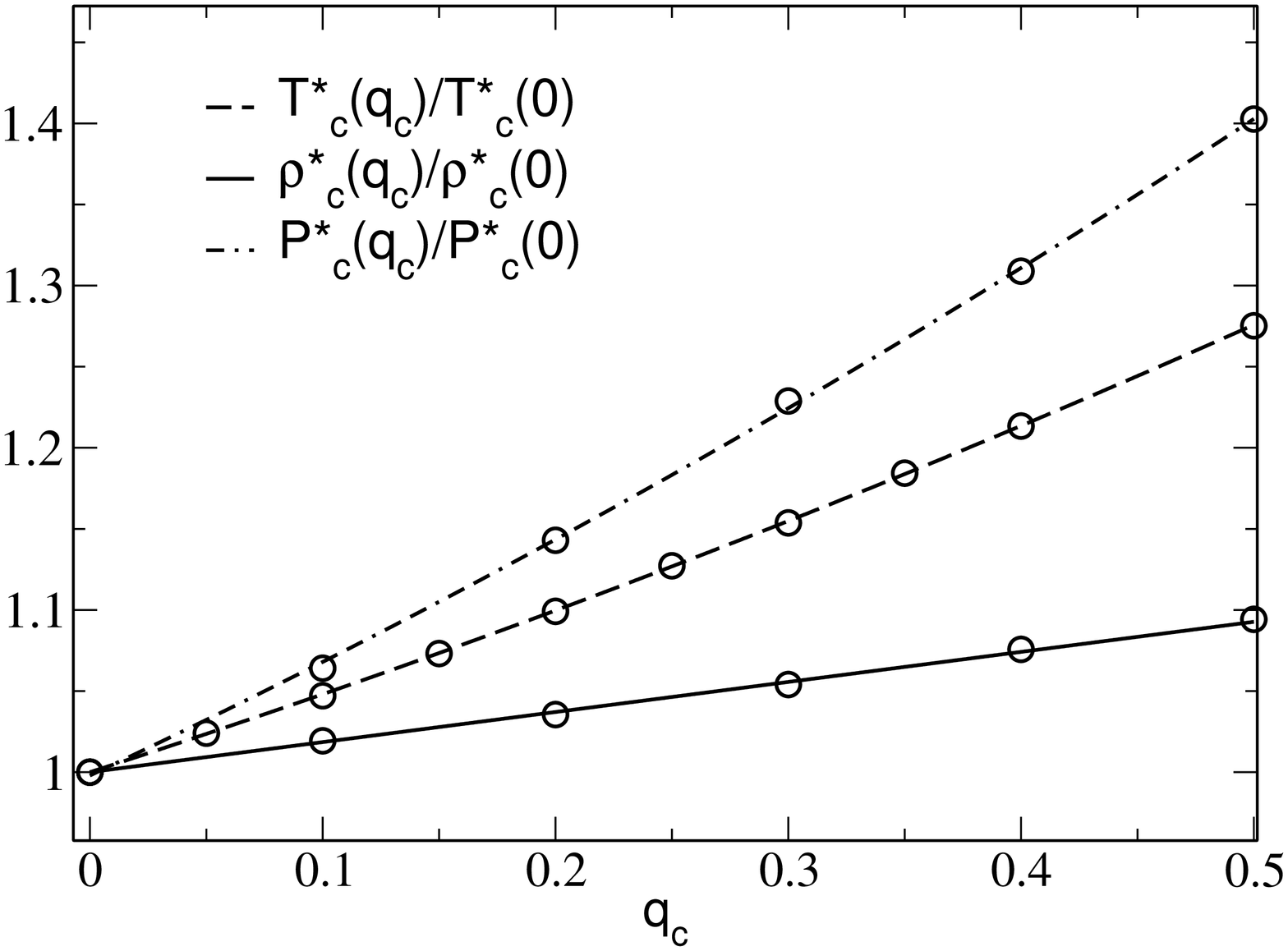}
\end{figure}
FIG.\ 1

\newpage
\clearpage
\begin{figure}
\includegraphics[angle=-90,scale=0.55]{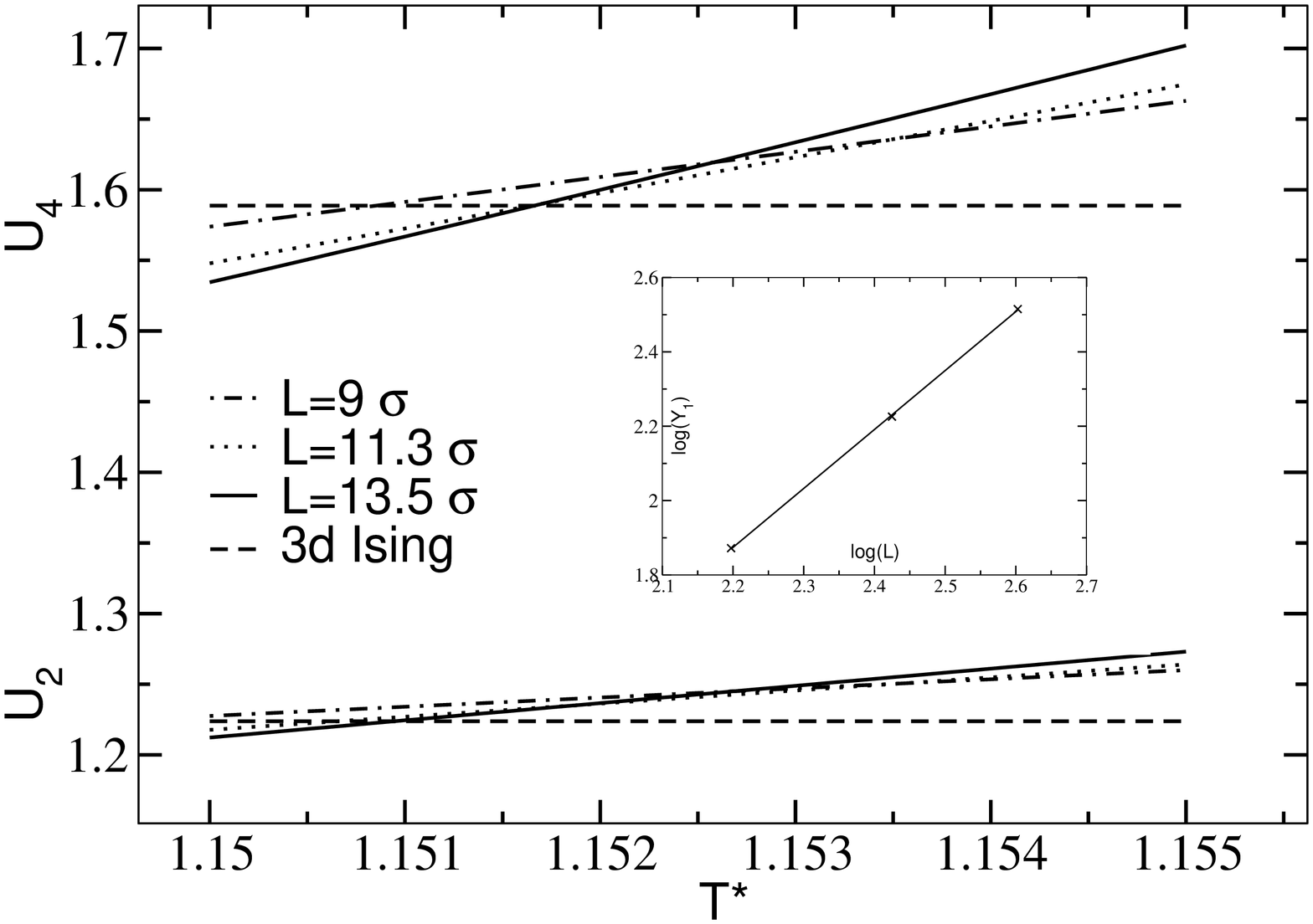}
\end{figure}
FIG.\ 2

\newpage
\clearpage

\begin{figure}
\includegraphics[angle=-90,scale=0.55]{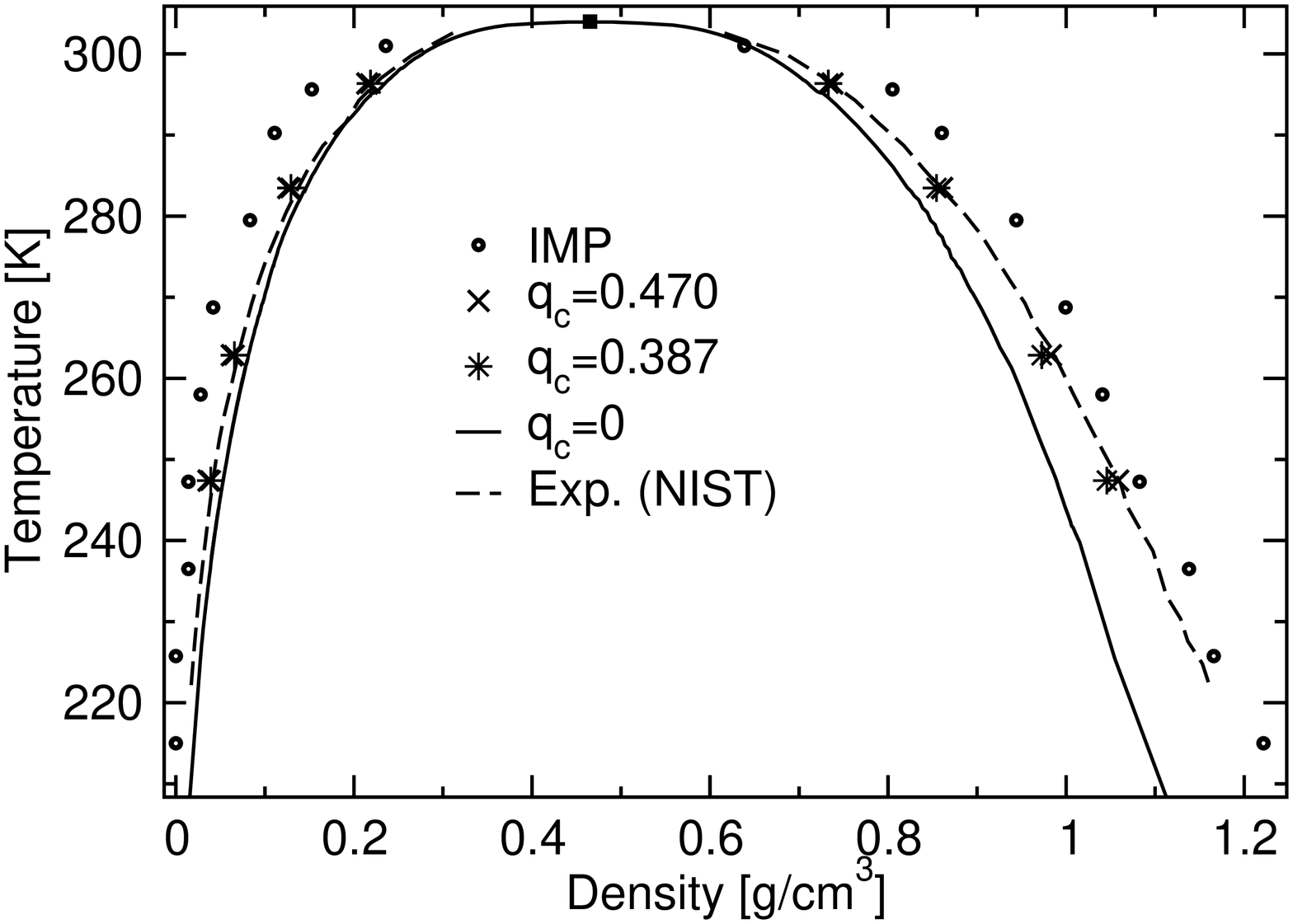}
\end{figure}
FIG.\ 3

\newpage
\clearpage

\begin{figure}
\includegraphics[angle=-90,scale=0.55]{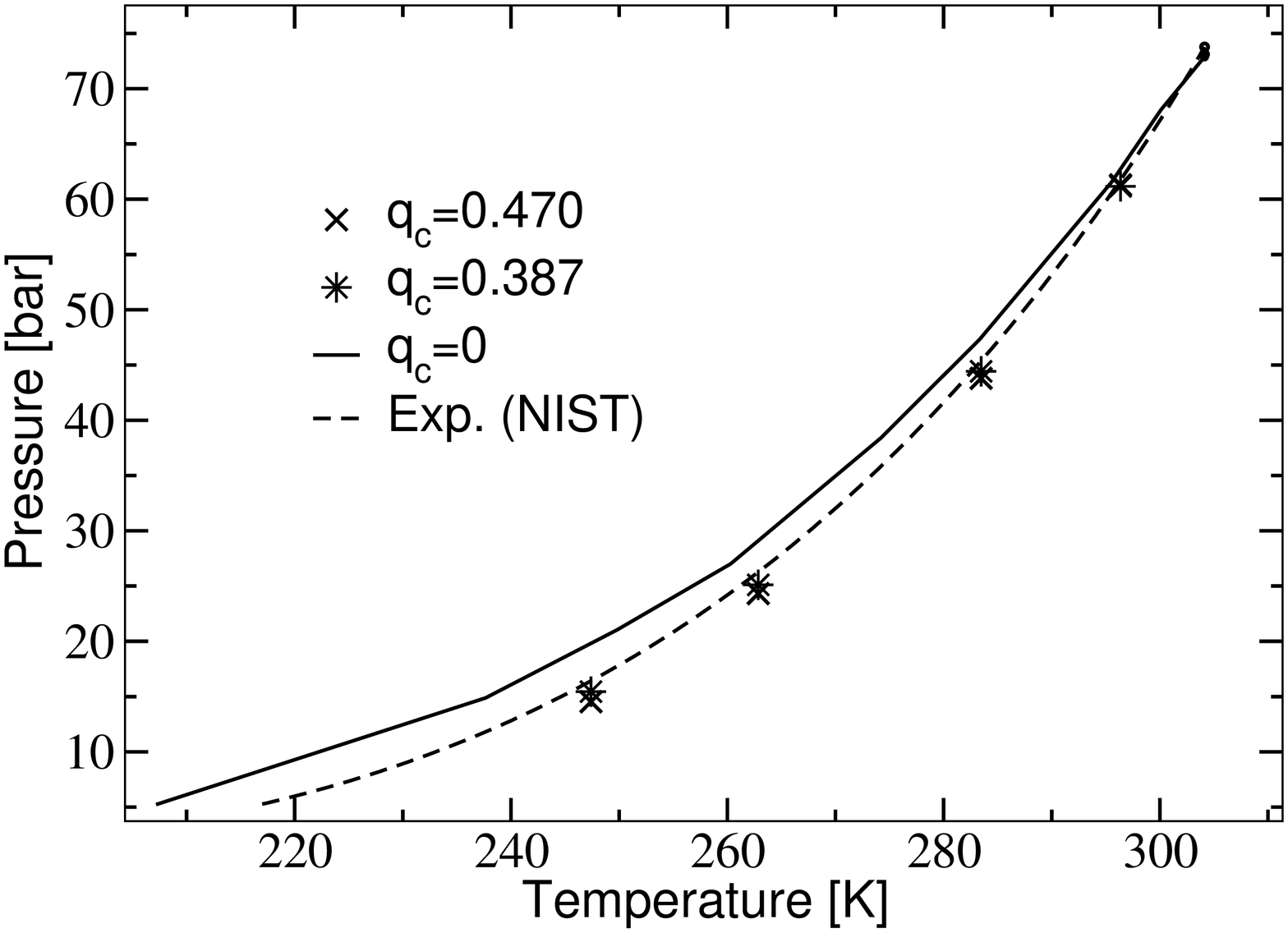}
\end{figure}
FIG.\ 4

\newpage
\clearpage

\begin{figure}
\includegraphics[angle=-90,scale=0.55]{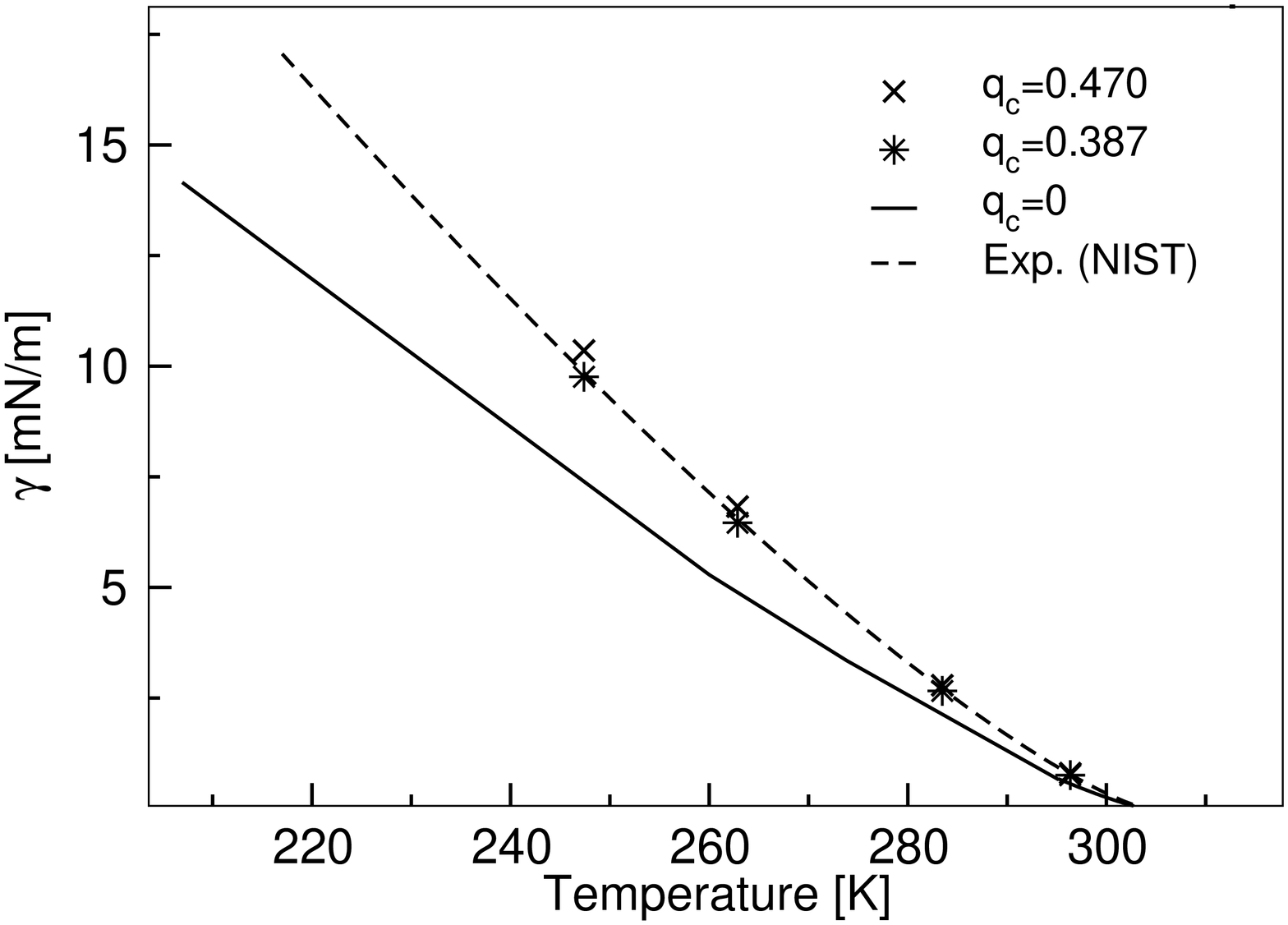}
\end{figure}
FIG.\ 5

\newpage
\clearpage

\begin{figure}
\includegraphics[angle=-90,scale=0.55]{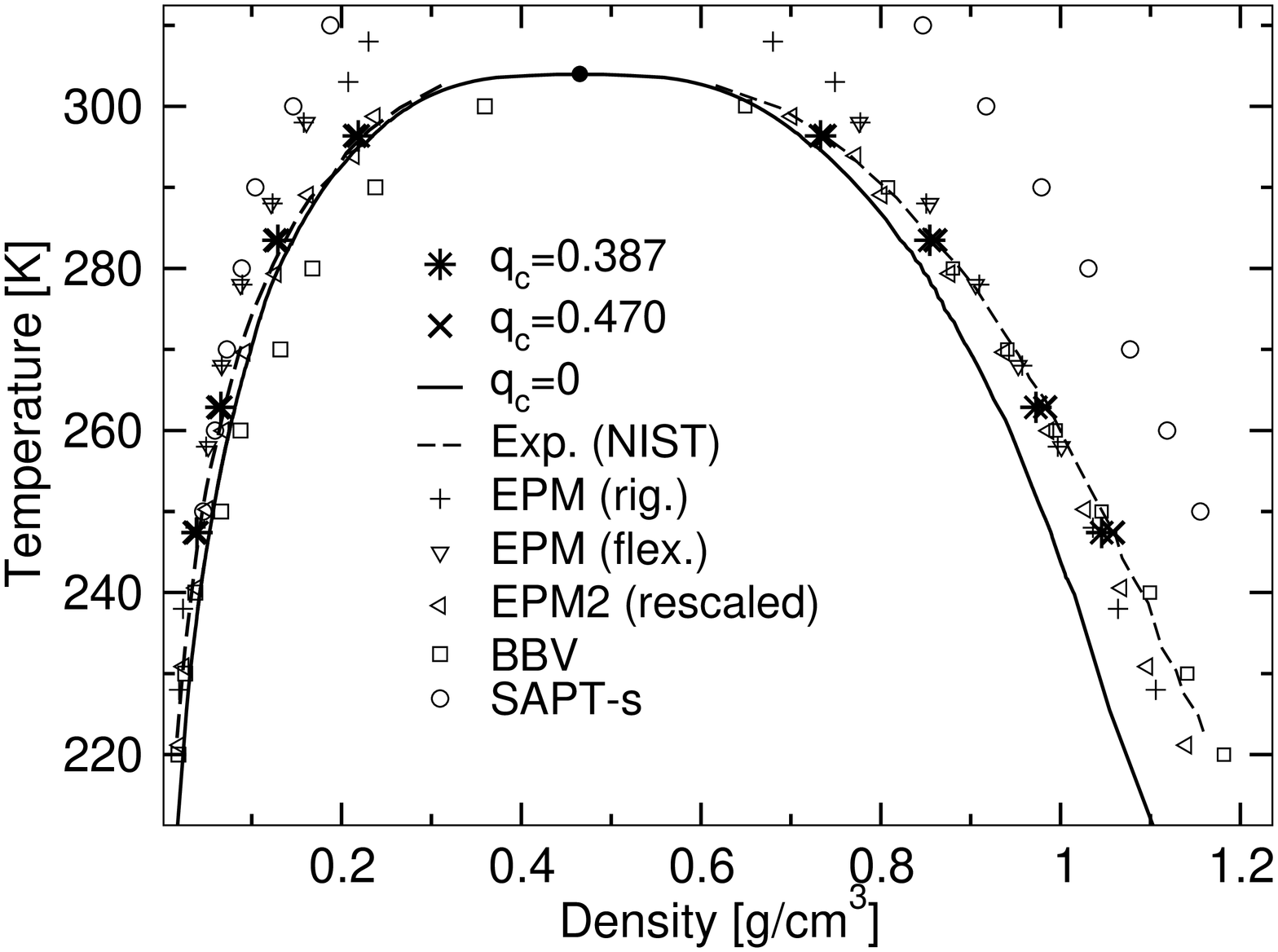}
\end{figure}
FIG.\ 6

\newpage
\clearpage

\begin{figure}
\includegraphics[angle=-90,scale=0.55]{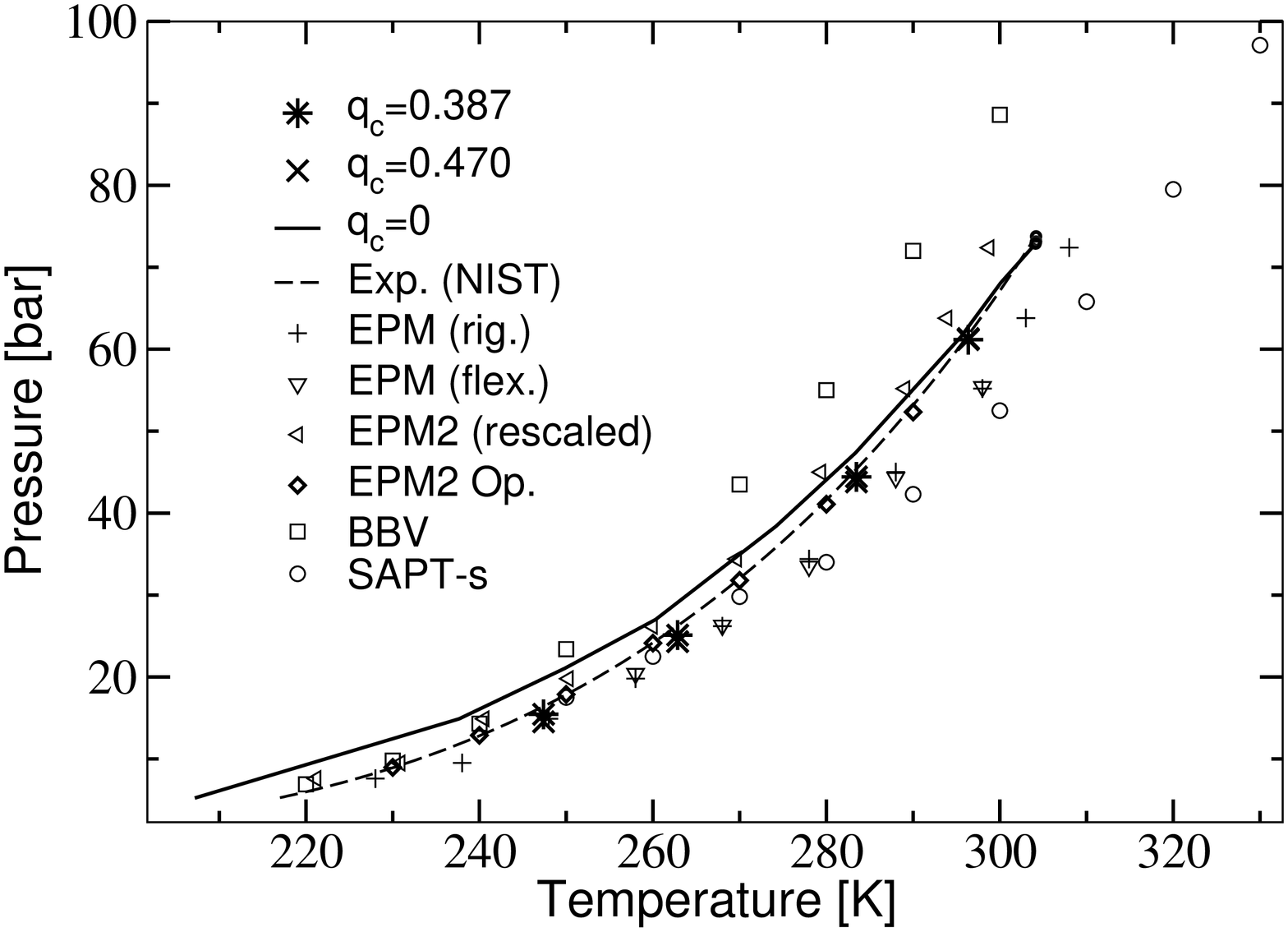}
\end{figure}
FIG.\ 7

\newpage
\clearpage

\begin{figure}
\includegraphics[angle=-90,scale=0.55]{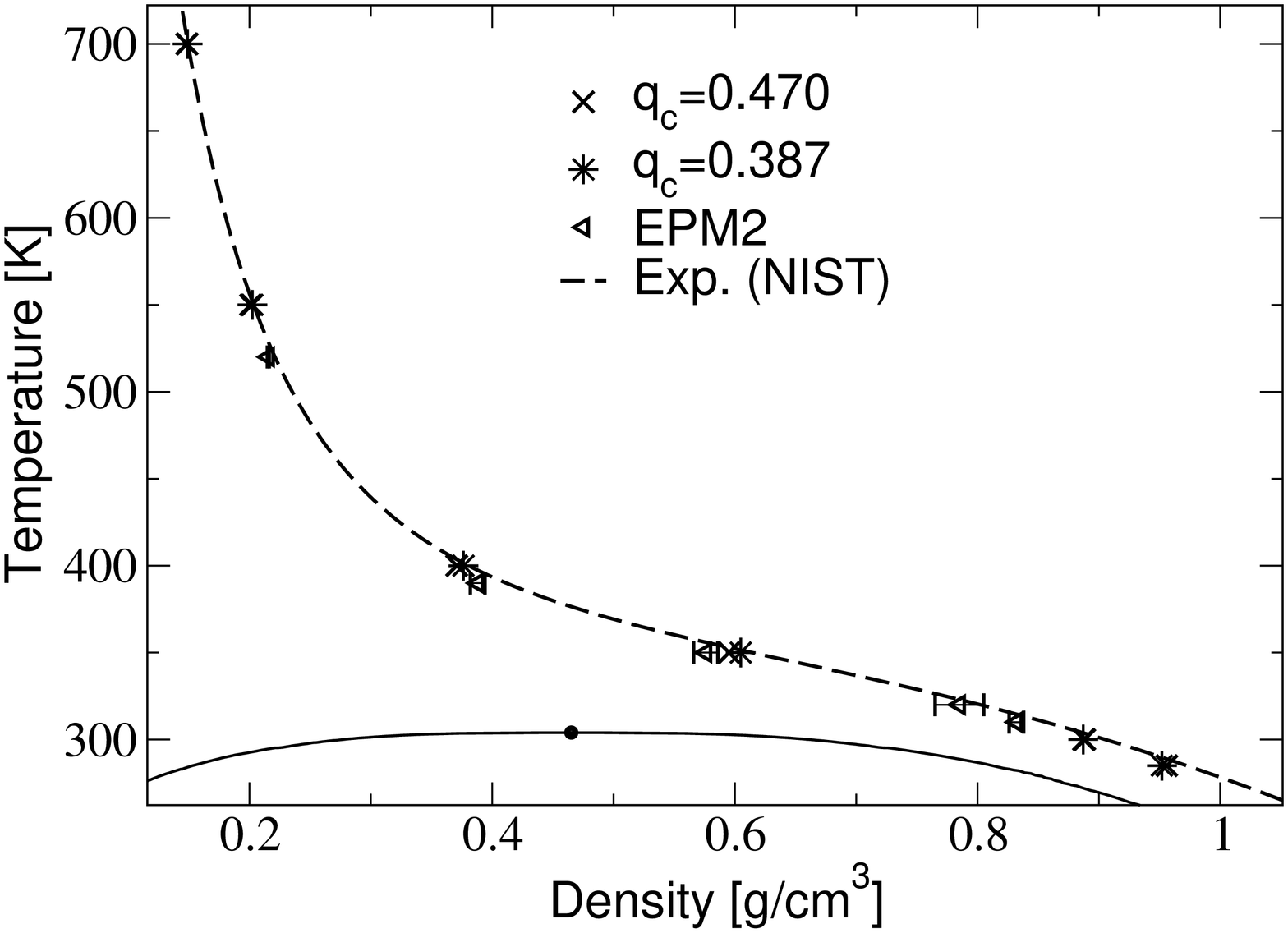}
\end{figure}
FIG.\ 8

\newpage
\clearpage

\begin{figure}
\includegraphics[angle=-90,scale=0.55]{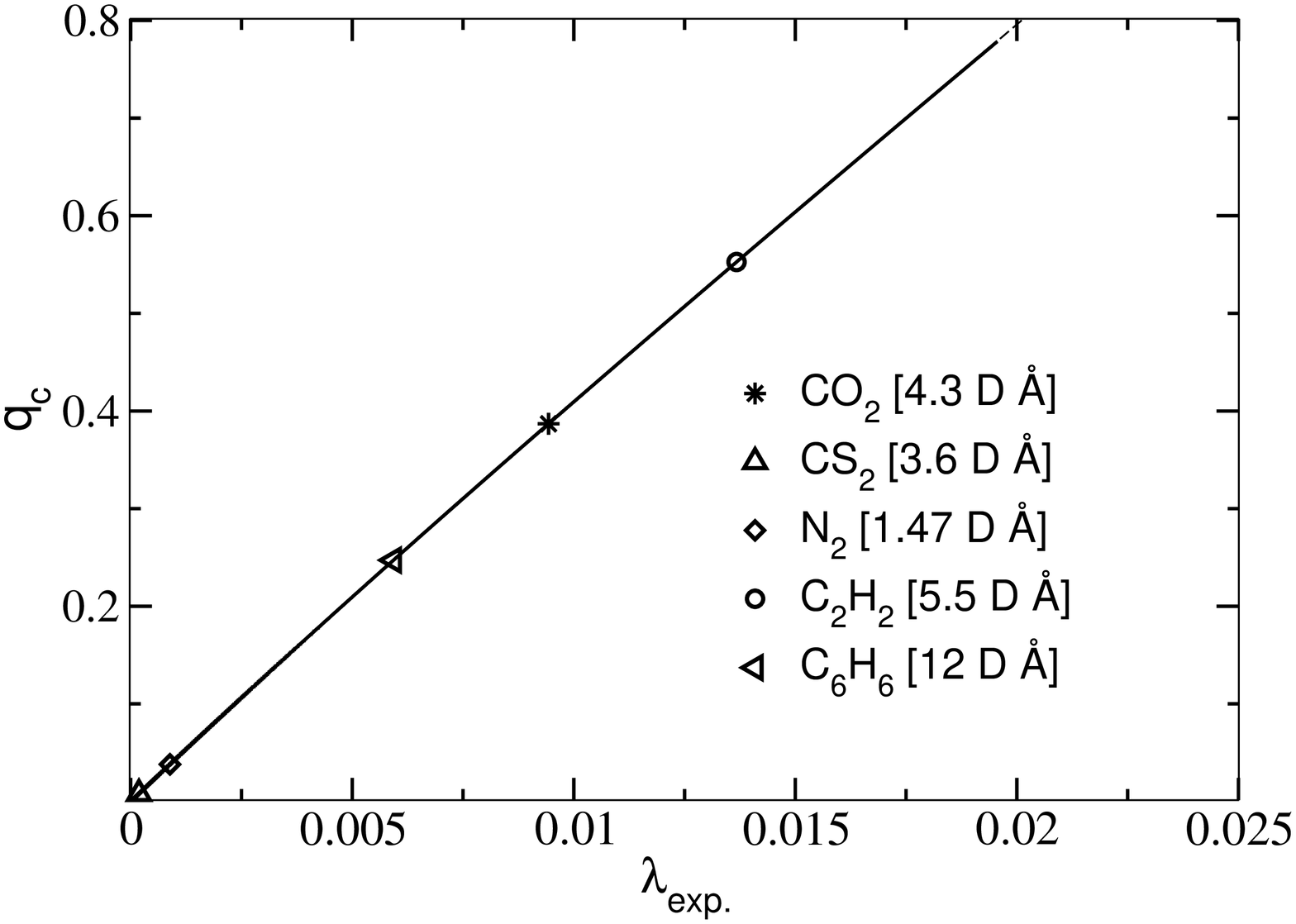}
\end{figure}
FIG.\ 9

\newpage
\clearpage

\begin{figure}
\includegraphics[scale=0.6]{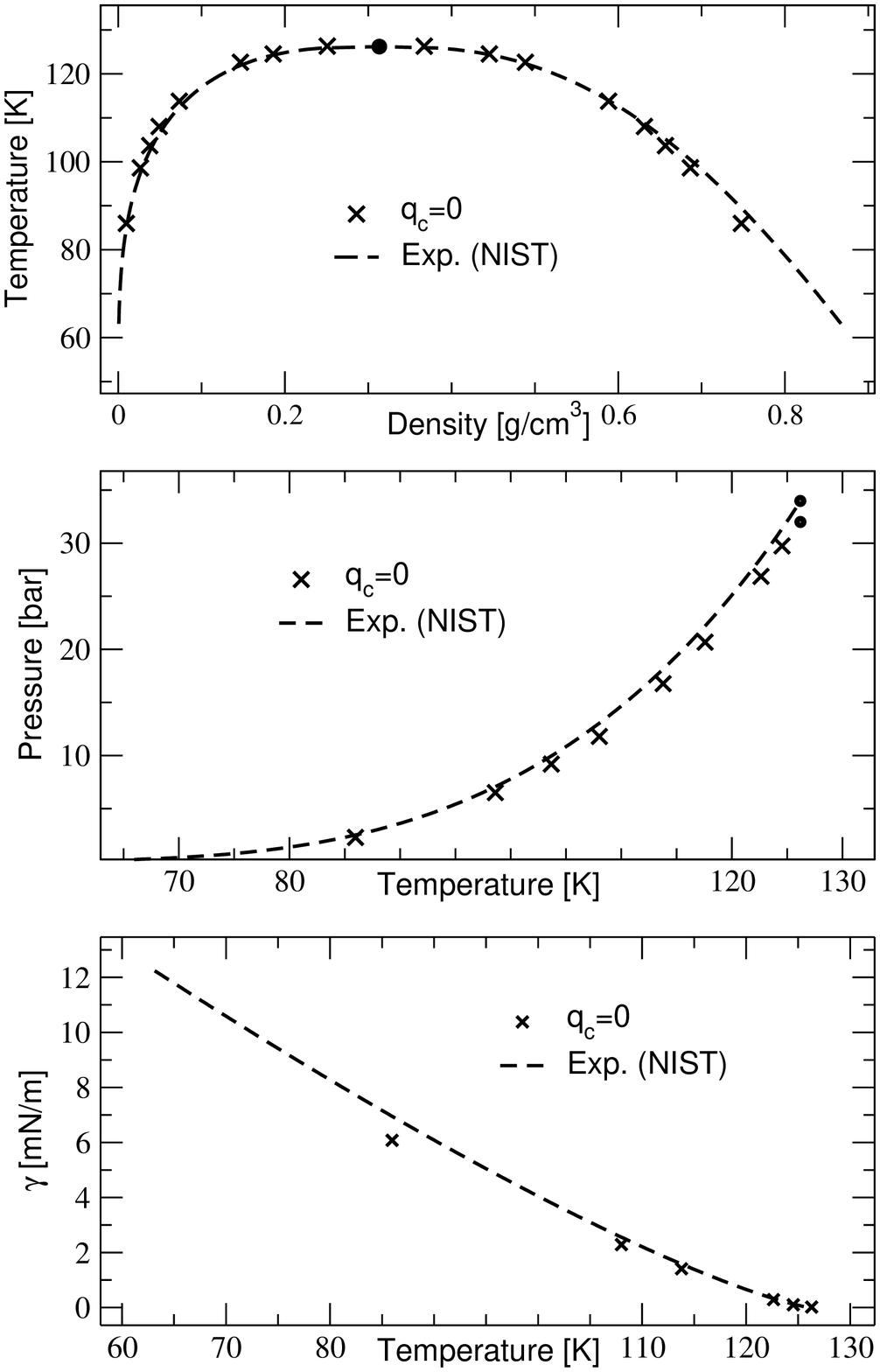}
\end{figure}
FIG.\ 10

\newpage
\clearpage

\begin{figure}
\includegraphics[angle=-90,scale=0.55]{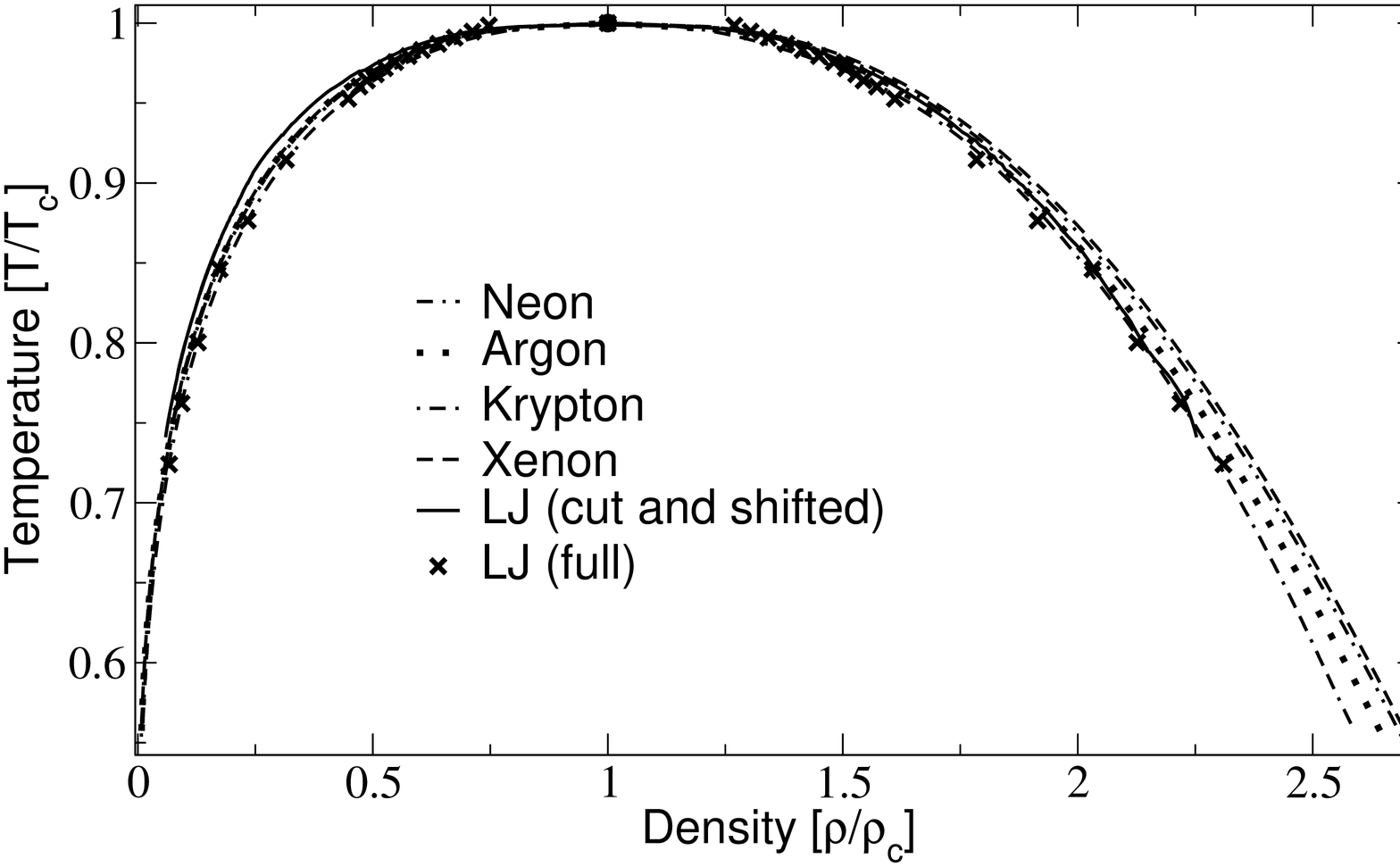}
\end{figure}
FIG.\ 11

\newpage
\clearpage

\begin{figure}
\includegraphics[scale=0.6]{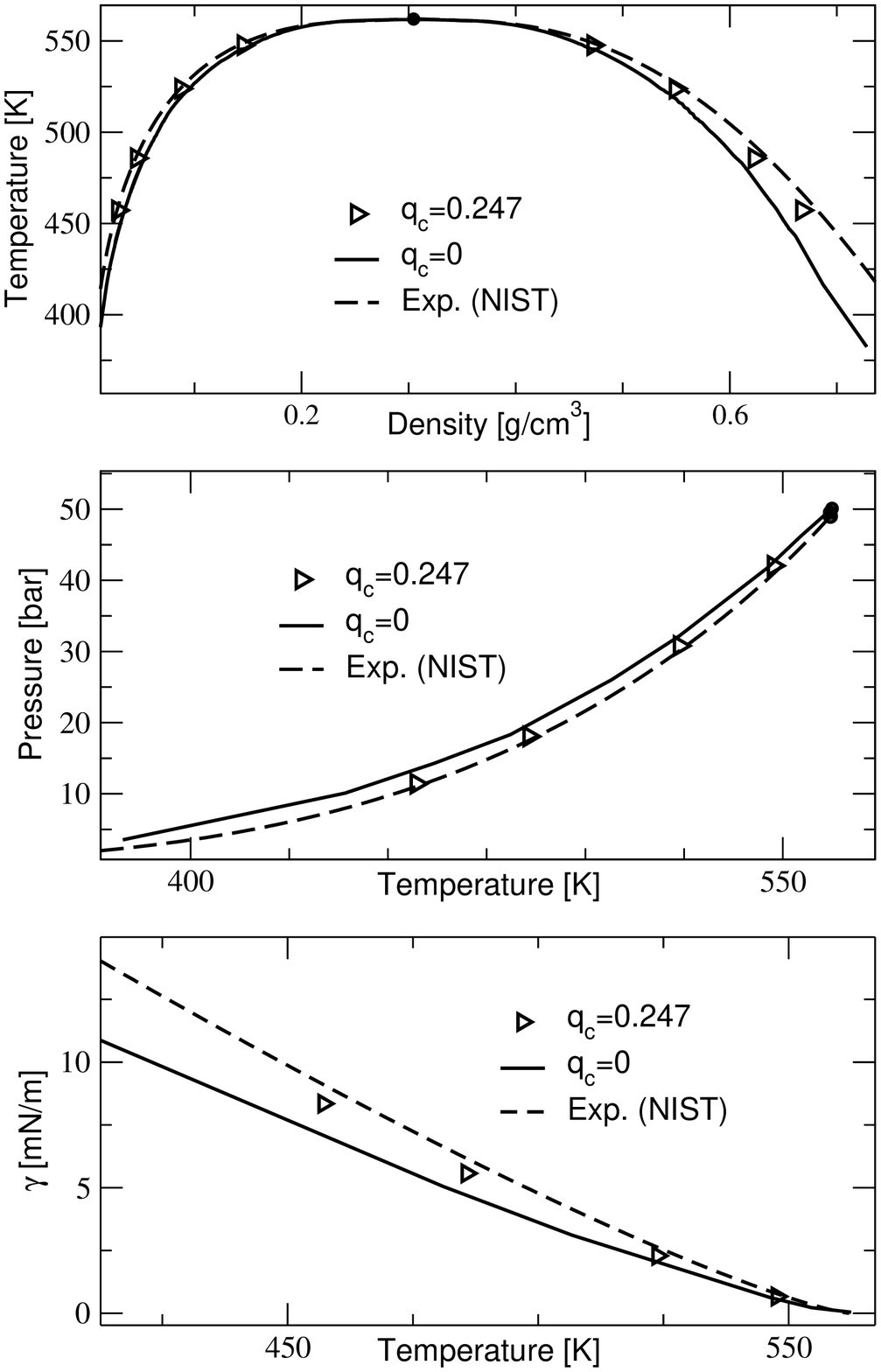}
\end{figure}
FIG.\ 12

\newpage
\clearpage

\begin{figure}
\includegraphics[scale=0.7]{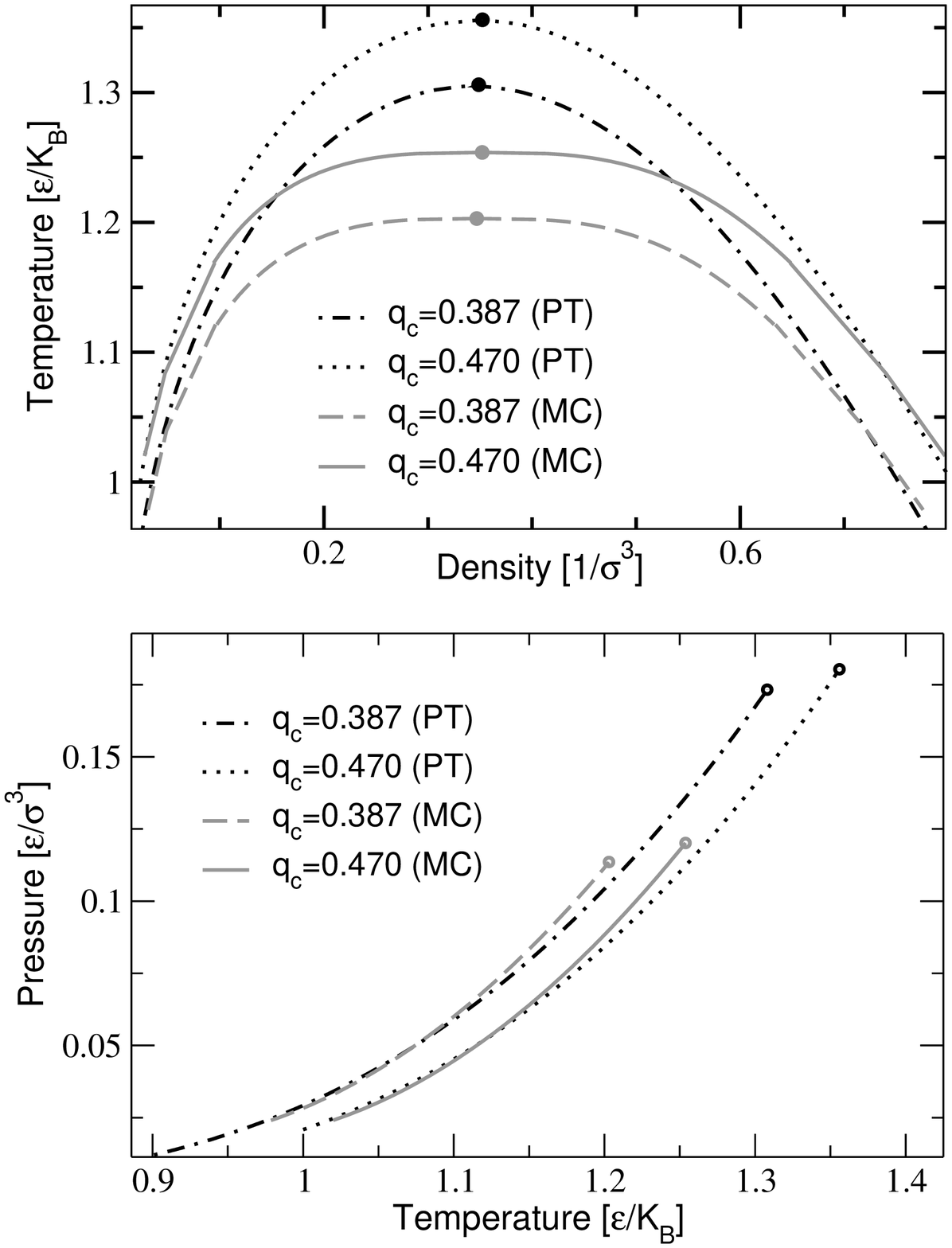}
\end{figure}
FIG.\ 13

\end{document}